\documentclass[a4paper,fleqn,usenatbib]{mnras}

\renewcommand{\footnoterule}{%
  \kern -19pt
  \hrule width 2in
  \kern 2.6pt
}
\usepackage{mathptmx}

\usepackage[T1]{fontenc}
\usepackage{ae,aecompl}

\usepackage{graphicx}	
\usepackage{amsmath}	
\usepackage{amssymb}	
\usepackage{scrextend}  
\usepackage{xcolor}     
\usepackage{subfig}     
\usepackage{longtable}

\title[SF in perturbed galaxies: effects of Tidal Interactions]{Star Formation in CALIFA survey perturbed galaxies.\\
I. Effects of Tidal Interactions}

\author[A. Morales-Vargas et al.]{
A. Morales-Vargas,$^{1}$\thanks{E-mail: abdmoralesv@gmail.com (AM)}
J. P. Torres-Papaqui,$^{1}$
F. F. Rosales-Ortega,$^{2}$
S. F. S\'{a}nchez,$^{3}$
\
\newauthor
M. Chow-Mart\'{i}nez,$^{4}$
R. A. Ortega-Minakata,$^{5}$
J. J. Trejo-Alonso,$^{6}$
\
\newauthor
A. C. Robleto-Or\'{u}s,$^{1}$
F. J. Romero-Cruz,$^{7}$
D. M. Neri-Larios$^{8}$
\
\newauthor
\& the CALIFA survey Collaboration$^{9}$
\\
$^{1}$Departamento de Astronom\'\i{}a, Universidad de Guanajuato, Apartado Postal 144, Guanajuato 36000, Mexico\\
$^{2}$Instituto Nacional de Astrof\'\i{}sica, \'{O}ptica y Electr\'{o}nica, Luis Enrique Erro 1, Tonantzintla 72840, Mexico\\
$^{3}$Instituto de Astronom\'\i{}a, Universidad Nacional Aut\'{o}noma de M\'exico (UNAM), Apartado Postal 70-264, M\'exico D. F. 04510, Mexico\\
$^{4}$Instituto de Geolog\'{i}a y Geof\'{i}sica, Universidad Nacional Aut\'{o}noma de Nicaragua, Rotonda Universitaria\\
$^{\phantom{0}}$Rigoberto L\'{o}pez P\'{e}rez 150 metros al Este, Managua 663, Nicaragua\\
$^{5}$Instituto de Radioastronom\'{i}a y Astrof\'{i}sica (IRyA), UNAM, Apartado Postal 72-3, Morelia, Michoac\'{a}n 58089, Mexico.\\
$^{6}$Facultad de Ingenier\'{i}a, Universidad Aut\'{o}noma de Quer\'{e}taro, Cerro de las Campanas s/n, Centro Universitario,\\ 
$^{\phantom{0}}$Santiago de Quer\'{e}taro 76010, Mexico\\
$^{7}$Instituto Tecnol\'{o}gico Superior de Guanajuato, Guanajuato 36262, Mexico\\
$^{8}$School of Physics, The University of Melbourne, Parkville, VIC. 3010, Australia\\
$^{9}$\texttt{https://califa.caha.es/}
}

\date{Accepted XXX. Received YYY; in original form ZZZ}

\pubyear{2020}

\begin{document}
\label{firstpage}
\pagerange{\pageref{firstpage}--\pageref{lastpage}}
\maketitle

\begin{abstract}

We explore the effects of tidal interactions on star formation (SF) by analysing a sample of CALIFA survey galaxies. The sample consists of tidally and 
non-tidally perturbed galaxies, paired at the closest stellar mass densities for the same galaxy type between subsamples. They are then compared, both 
on the resolved Star Formation Main Sequence (SFMS) plane and in annular property profiles. Star-forming regions in tidally perturbed galaxies exhibit 
flatter SFMS slopes compared to star-forming regions in non-tidally perturbed galaxies. Despite that the annular profiles show star-forming regions in 
tidally perturbed galaxies as being mostly older, their SF properties are never reduced against those ones proper of non-tidally perturbed galaxies. 
Star-forming regions in non-tidally perturbed galaxies are better candidates for SF suppression (quenching). The lowered SF with increasing stellar mass 
density in tidally perturbed galaxies may suggest a lower dependence of SF on stellar mass. Though the SFMS slopes, either flatter or steeper, are found 
independent of stellar mass density, the effect of global stellar mass can not be ignored when distinguishing among galaxy types. Since a phenomenon or 
property other than local/global stellar mass may be taking part in the modulation of SF, the integrated SF properties are related to the tidal perturbation 
parameter. We find weak, but detectable, positive correlations for perturbed galaxies suggesting that tidal perturbations induced by close companions 
increase the gas accretion rates of these objects.

\end{abstract}




\section{Introduction}
\label{sec:int}

Characterizing the unlike or opposite natures of \emph{passive} and \emph{forced} galaxy evolution via \emph{quiescent} and \textit{induced} star formation (SF) has 
been vastly intriguing. Interactions are, undoubtedly, typical \emph{actuators} of SF. In galaxy \emph{mergers} and \emph{pairs}, gravitational tidal effects use to 
induce SF by overruning the self-gravity of the progenitors \citep[\textit{e.g.}][]{BarHer91,BarHer96,WoGe07}. 

In galaxy mergers, enhanced conversions of both molecular and atomic gas may yield SF efficiencies of at least one order of magnitude larger \citep[\textit{e.g.}]
[]{MiRiBo92,BeKo94,MiHe94,MiHe96,Yo99,Li08}. For these cases, simulating the response of global SF is complex: it depends on orbital dynamics, aligned disk spin 
orientations of the progenitors, their gas fraction and distribution, mass ratios \citep[\textit{e.g.}][]{MiRiBo92,MiHe96,Tis02,BLA03,Cox04,Pe05,Dav15}; as well 
as models for prescribing SF and feedback \citep[\textit{e.g.}][]{Spr00,Bar04,Spr05,Hop13}. Whereas the lower-mass (secondary) galaxy in \textit{minor} mergers 
appears to be the most affected by the interaction (\textit{e.g.} \citealt{Al12} and references therein), both galaxies in \textit{major} mergers use to suffer of 
enhanced SF \citep[\textit{e.g.}][]{Mast05,WoGe07,Dav15,Mor15}. Major mergers are relatively easy to identify whereas minor mergers, more frequent in the local 
Universe, may also contribute to drive galaxy evolution (\citealt{Ven19} and references therein). 

Though galaxy pairs also show molecular gas fraction enhancements due to tidal torques \citep[\textit{e.g.}][]{Vio18}, passing-by encounters are less 
effective in triggering SF \citep[\textit{e.g.}][]{MiRiBo92,WoGe07}. However, retrograde encounters of the flyby-passing disks may increase such that efectiveness 
\citep[\textit{e.g.}][]{Wil14}. Not least in importance, the merger fraction and in general the build up of stellar mass greatly depend on the estimation of 
galaxy pairs \citep[\textit{e.g.}][]{Ya-Ch03,Kee13,Ven19}. 

Either hiked or weakly raised, most induced SF is centrally located \citep[\textit{e.g.}][]{HeMi95,MiHe94,MiHe96,Spr00,May01,Yu12,Hop13,Mor15,ArFe16}. This may be 
due to gas inflows \citep[\textit{e.g.}][]{Cap15,Ble18} though important amounts of gas are either ejected by winds \citep[\textit{e.g.}][]{Hop08,Wil14,LopC17,LopC18} 
or stripped off from galaxies \citep[\textit{e.g.}][]{DiM08,Bit10}. Since off-central SF is either hard to trigger or so short-lived, interactions prevail as activators 
of gas inflows. 

Additionally, galaxy environment peculiarly affects SF. Measurements of compaction/expansion of a collection of objects within a certain limited phase space have 
been done so far \citep[\textit{e.g.}][]{Lew02,Gom03,Kauff04,Gav10,Cal11,Vul15,Sch17}. Recently, \citet{Zhe17} model the morphology of the local cosmic density 
field. \citet{God17} split into centrals and satellites to compare Stellar Population (SP) gradients. \citet{Gav10} and \citet{God17} particularly discuss the 
biases a local density parameter may yield. If appearing far from a main agregate (\textit{i.e.}, being an outlier in the velocity distribution) the density of 
an object would be lower than the true one, that if there were indeed significant gravitational interactions in the aggregate. If background/foreground objects 
not physically related to the object and aggregate were present, evaluating the density of the former would be senseless.

Another concept in line with galaxy evolution is the so-called Star Formation Main Sequence \citep[SFMS, \textit{e.g.}][]{Brinch04,Elb07,Salim07,Pen10,Whi12,Spe14,
RenPen15,Can16,Er16,Cat17,Ell18a,Lop18,San18a,Med18,EF19}. From internal to external processes, it has been assisting to identify what modulates SF, specially, up 
to $z\,\sim1$. Great debate however has emerged due to either its uncertainties \citep[\textit{}e.g.][]{Elb07,Pen10}, or the fact that it should not be considered 
a linear relation \citep{Er16}, mainly, at log$_{10}$ M$_{*}\,$(M$_{\odot}$)\,$>\,$10. Certainly, a flat slope characterizes the SFMS for these masses \citep{Whi12,
Er16}, often related to bulge-dominated galaxies, what gives the sequence a large dispersion \citep{Schi07,Salim07,Whi12,Gon16}. A quite broad sequence for late 
type galaxies has even been reported due to the stochasticity of the star-forming regions \citep{Vul19}.

If adapting a linear model, the SFMS logarithmic slope has resulted quite fluctuating \citep[\textit{e.g.}][]{Elb07,Spe14,RenPen15,Can16,Mar17,San18a}. \citet{{Spe14}} 
show this is due to its dependence on time evolution and other not less important factors (Initial Mass Function, IMF; Star Formation Rate SFR tracers; SP models; 
etc.). By confirming this time dependency, \citet{Lop18} and \citet{San18a} have unveiled the cosmic SF quenching not as simple as a one-event process since a fraction 
of high-$z$ passive galaxies have become rejuvenated at lower redshifts and ended quenched later on. 

By featuring the SFR intensity ($\mathrm{\Sigma_{SFR}}$) and stellar mass surface density ($\Sigma_{*}$) in a spatially-resolved SFMS, \citet{Can16} and \citet{Gon16} 
predict a slightly steeper integrated (global) sequence. This appears to have its origin in the spatially-resolved sequence \citep{Hsi17,Mar17,Can19}. Generally, the resolved SFMS 
assists to figure out how complex the regulation of SF is \citep[external, global and local actuators,][]{Med18}. For instance, by integrating galaxy components, 
\citet{Cat17} show massive disks as having undergone efficient SF suppression. \citet{Hsi17} find reduced fractions of H \textsc{ii} regions from the periphery of 
quenched galaxies. Later, from SFMS offsets, \citet{Ell18a} show that SF enhancements/suppressions occur inside-out. \citet{Hal18} find sequences with contrasting 
patterns which may result from the rate of mass inflows. \citet{Can19} propose that local SF is indirectly modulated by galaxy morphology.

A pair of goals are introduced with all this background. To get insight on centrally driven/located gas/SF and their plausible relation with tidal interactions 
(\textit{e.g.} \citealt{Ell18b} and references therein), we compare the $\mathrm{\Sigma_{SFR}}$ (SFR\,kpc$^{-2}$) annular profiles of star-forming regions within 
tidally and non-tidally perturbed galaxies. Instead of a local density measurement, we treat each tidally perturbed object by simply considering its closest neighbour. 
There is no distinction, for instance, if centrals or satellites, but just galaxies under tidal torques, neither in rigorous established pairs nor in groups nor in 
clusters. The quite challenging task of establishing a general characterization of environment justifies this approach since different estimations are relevant for 
different physical effects \citep{Wal14}.

Secondly, as close encounters use to unbalance SF, we look for a possible dependence of the resolved SFMS on unlike degrees of interaction. To do so, the 
star-forming regions within our non-tidally/tidally perturbed galaxies are pictured in the $\mathrm{\Sigma_{SFR}}$-$\Sigma_{*}$ plane.

For a good direction of both goals we use:

\begin{enumerate}
 \item Integral Field Spectroscopy (IFS), perfectly suitable to spatially split up detailed distributions of any property of concern. The Calar Alto Legacy Integral 
 Field  Area \citep[CALIFA,][]{San12a,Hus13,Gar15,San16a} survey is used for this purpose. The CALIFA survey favorably presents the best compromise among near-by  
 coverage ($0.005\leq\,z\,\leq0.03$), spatial coverage (mostly to 2.5 effectuve radius), spatial resolution ($\sim1\,$kpc), number of targets ($>600$) and target 
 sampling ($\sim4,000$ spectra per target). \item Synthesis of Stellar Populations (SSPs) by applying detailed spectral synthesis techniques \citep[\textit{e.g.}]
 []{Cid05,Asa07}.
\end{enumerate}

This paper is ordered as follows. Methods to obtain the SP properties are described in Section~\ref{sec:met}. Our samples are defined in Section~\ref{sec:sam}. We 
present resolved SFMSs and our property profiles in Section~\ref{sec:res}. We discuss both results in Section~\ref{sec:dis}. Summary and conclusions are stated in 
Section \ref{sec:conc}.

We use a cosmological set of $\mathrm{H_{0}}\,=72\,\mathrm{h_{72}^{-1}}\,\mathrm{km\,s^{-1}\,Mpc^{-1}}$, $\Omega_\mathrm{M}\,=\,0.3$ and $\Omega_\mathrm{\Lambda}\,=\,0.7$; 
a \citet{Chab03} IMF for SFR and stellar mass (M$_{*}$) estimations; and a 0.05 level of significance for all statistics.

\section{Methods}
\label{sec:met}

\subsection{Stellar component subtraction and emission line fitting}
\label{subsec:continuum_line}

One spectrum is contained in the third (wavelength) dimension of each spaxel\footnote{An IFS discrete spatial element \citep{Ros10}.}. These are extracted, read, 
and selected only those with at least one non-NULL value (typically $\sim$4\,000, \textit{i.e.}, $\sim$78\,\% of a data cube\footnote{This reflects the unavoidable 
fiber loss of throughput close to the edges and gradually increasing towards the corners of the instrument detector \citep{San12a}.}). Our processing pipeline 
rebins this selection to the resolution of the \textsc{starlight} code \citep[SSPs,][]{Cid05}. The code version relies on the MILES base of spectral libraries 
\citep{SaB06,Fac11} and uses the simple SPs of \citet{BrCh03} synthesis models (\citealt{Chab03} IMF). \textsc{starlight} satisfactorily solves spectra with no 
NULLs along the wavelength dimension (typically $\sim$3\,000, \textit{i.e.}, $\sim$60\,\% of a data cube). The nearly pure nebular spectra, result from subtracting 
the stellar syntheses, are taken to fit the emission lines of interest by adapting Gaussian profiles. Central wavelength, amplitude and associated dispersion for 
each line are initial parameters. Iterations are done till finding the minimum $\chi^{2}$ value (residual) between the observed line and the best profile. Isolated 
lines are fit individually whereas multiple profiles ($G=G_{1}+G_{2}+...+G_{n}$) are constructed for blended lines. The signal-to-noise (S/N) at the observed central 
wavelength of each emission line serves to estimate flux uncertainties. Full width at half maximum (FWHM) and wavelength displacements are also estimated. 

\subsection{Galaxy morphologies and colours} 
\label{subsec:gal-col}

The CALIFA survey Collaboration (hereafter ``the Collaboration'') carried out a morphological re-classification for all galaxies of the CALIFA survey Mother Sample 
(MS, \textit{i.e.}, the set of candidates for the survey observations, see \citealt{Wal14}, W-14 from now on). On Sloan Digital Sky Survey (SDSS) images (\textit{r} 
and \textit{i} bands), five collaborators based their respective visual classifications on the following: 1) E, S, or I for elliptical, spiral or irregular; 2) 0-7 
for E; 0, 0a, a, ab, b, bc, c, cd, d and m for S; or r for I; 3) B for barred, A for unbarred or AB if unsure; and 4) merger features, yes (Y) or no (N). The five 
classifications were combined to obtain each mean by ignoring outliers. Appendix \ref{sec:app1} lists the resultant morphologies for the galaxies involved in this 
work.

Galaxy colours are determined from colour magnitude diagrams (CMDs) which use SDSS/DR7 \citep{Aba09} model magnitudes\footnote{Model magnitudes are optimal measures of 
fluxes of galaxies. They result from fitting two galaxy models on each object in each band. The highest likelihood model in the \textit{r} band (modelMag, 
https://bit.ly/3e4wKm5) is chosen and applied to the other bands after convolving with the point spread function in each band.} (see Fig.~\ref{f3}). 
The Eqs. giving the cuts to select ``red'' and ``blue'' galaxies are:
\setlength{\mathindent}{0cm}
\begin{equation} \label{eq:1}
 \begin{split}
g-r\,=\,-0.0371 \times (\mathrm{M}_{r} + 24) + 0.81 \;\;\mathrm{and}\\
g-r\,=\,(-0.0371 \times (\mathrm{M}_{r} + 24) + 0.81)-0.12,
 \end{split} 
\bigskip
\end{equation}

\noindent where M$_{r}$ is the \textit{r}-band absolute magnitude. Both Eqs. are within a 0.98 confidence interval and represent correlations for the ``red 
sequence'' and ``blue cloud'' respectively. We derive them by using Eqs. 1 and 2 of \citet{Schaw14} on data of all SDSS/DR7 objects. Finally, ``green'' galaxies are 
in-between cuts. 

\subsection{Star-forming regions, SFRs and stellar masses} 
\label{subsec:s-f_r}

\subsubsection{Star-forming regions, SFR \& $\varSigma_{SFR}$ estimations}
\label{subsec:s-f_rsub}
Prior to define the star-forming regions, the dominant source of gas excitation is determined. The \citet{BPT81} diagram (BPT, their fig. 5) is the standard tool 
for this. Line demarcations used for pure star-forming galaxy (SFG) and active galactic nucleus (AGN) excitations are respectively those of \citet{Kauff03} and \citet{Kew01}. In-between excitation is often dubbed 
as Transition Object (TO). \citet{Tor12b} demarcate Seyfert 2 (Sy2) excitation and Low Ionization (Nuclear) Emission line Regions (LI(N)ERs). 

Our pipeline for analysis applies then the SFG spectral characterization of \citet{Cid07,Cid10} and \citet{Asa07}. It requires that the four emission lines that 
construct the BPT diagram fulfill the line criteria of Table~\ref{tab:1} (first-row). If a ``resolved'' BPT diagram can be extracted from a galaxy, this will be 
an Emission Line Galaxy (ELG) since both the H$\alpha$ and N\,[II]$\lambda$6583 lines have a S/N\,$\geq$3 \citep{Cid10,Tor12a}. It is implicit, later in the text, 
the preference given to objects containing star-forming regions. Active objects like these are the cornerstone to portray galaxy evolution in terms of SP 
properties. To assign the dominant excitation source (see Table~\ref{tab:A1}), the comparison of line ratios is done by previously integrating (summing up) all 
resolved fluxes of each involved line.

For star-forming regions, we proceed as follows. For all galaxy sets, the pipeline for analysis selects spectra which pass the full line criteria for H$\alpha$, and 
only the flux criterion for the rest lines of the BPT. This is due to two facts: 1) the H$\alpha$ line emission is our SFR tracer, and 2) if all line criteria is 
applied on the rest lines, the [\ion{N}{ii}] one, mainly, reduces the number of star-forming regions, usually, in blue galaxies. Next, we truncate each set by using 
an EW cut-off. Besides of proving strong excitation, an EW (H$\alpha$) cut-off of $\geq$\,6 \AA{} characterizes both, star-forming regions with [\ion{O}{ii}]-[\ion{O}{iii}] 
line S/N $\geq$3 \citep{Cid10}, and \ion{H}{ii} regions with big fractions of young SPs \citep{San14}. This truncation by itself effectively omits spaxels of observation 
artefacts and those of foreground stars.

\begin{table}
 \setlength{\tabcolsep}{0.5\tabcolsep}
\begin{minipage}{\columnwidth}
\centering
 \caption{\scriptsize{Emission line and mass-and-age criteria \citep[\textit{e.g.}][]{Cid07,Cid10,Asa07}.}
  \label{tab:1}}
 \begin{scriptsize}
 \begin{tabular}{lllll}
 \hline
\textsc{starlight} S/N                                                                                               &                                               &Observed                &                                                                                                                                             &Line                          \\
(continuum                                                                                                           &Emission                                       &emission                &Emission                                                                                                                                     &displacement                  \\
window:                                                                                                              &line flux ($\times\,10^{-16}$                  &line $\sigma$           &line                                                                                                                                         &$\lambda_{lab}-\lambda_{obs}$ \\
5075--5125 \AA{})                                                                                                    &$\mathrm{erg\,s^{-1}\,cm^{-2}}$\,\AA{}$^{-1}$) &($\mathrm{km\,s^{-1}}$) &S/N                                                                                                                                          &(\AA{})                       \\
\hline
                        $\geq5$\footnote{\scriptsize{Excitation sources and SFRs (Sections~\ref{subsec:s-f_rsub}).}} &1000$\geq$\,flux\,$\geq0.1$                    &$<400$                  &$\geq3$\footnote{\scriptsize{Appropriate lower limit for H$\beta$ and [O III]$\lambda$5007 line detections \citep[\textit{e.g.}][]{Cid10}.}} &$<10$                         \\
$\geq10$\footnote{\scriptsize{Approximation of global stellar mass and SP median age (Section~\ref{subsec:M_est}).}} &$\ldots$                                       &$\ldots$                &$\ldots$                                                                                                                                     &$\ldots$                      \\
\hline\\
 \end{tabular} 
 \end{scriptsize}
 \end{minipage}
\end{table} 

For the extinction of H$\alpha$ flux, the pipeline re-iterates, in each truncated set, the full line criteria on now the H$\beta$ line. Such that line mostly succeeds 
the criteria. For just several galaxies, failed spectra are a $\sim$5\,\% or less. Extinction correction is not applied in these failed cases. We use an intrinsic 
Balmer ratio of 2.86 for Case B recombination at T$_{e}=10,000\,$K and n$_{e}=100\,\mathrm{cm^{-3}}$ \citep{Hum87}. We use equation 1 of \citet{Cat15}:
\begin{equation} \label{eq:2}
A \left( H\alpha \right) = \left[ \frac{K_{H\alpha}}{-0.40 \left( K_{H\alpha} - K_{H\beta} \right)} \right] \cdot \log_{10}{\left[ \frac{F_{H\alpha}/F_{H\beta}}{2.86} \right]},
\bigskip
\end{equation}

\noindent where $K_{H\alpha}=2.54$ and $K_{H\beta}=3.61$ are the extinction coefficients from the Galaxy extinction curve \citep{Car89}. If $F_{H\alpha}/F_{H\beta}<$2.86, 
no extinction is assumed. The pipeline then takes each EW-truncated, extinction-corrected set to assign excitation sources to each single region. Only those with SFG 
excitation are selected for the conversion of \citet{Asa07}:
\begin{equation} \label{eq:3}
\mathrm{SFR}\,(\mathrm{M_{\odot}\,yr^{-1}})=5.2\times10^{-42}\,L_{H\alpha}\,(\mathrm{erg\,s^{-1}}),
\bigskip
\end{equation}

\noindent which uses a \citet{Chab03} IMF ($0.1$--$100\,\mathrm{M_{\odot}}$) to ensure the most ionizing stars and a SF constancy of the order of their lifetime 
($\sim$10\,Myr). Notice then that, regardless of the dominant excitation source determined earlier, the star-forming regions are defined as those spaxels whose 
spectra show EW (H$\alpha$)\,$\geq$\,6 \AA{}\footnote{Mostly compact regions. \citealt{Lac18} prove that an EW (H$\alpha$)\,$>\sim\,$10\,\AA{} distinguishes 
star-forming from diffuse ionized gas regions.} and that lay below the \citet{Kauff03} demarcation in the BPT.

Global SFRs are the sum of SFRs of all star-forming regions. These resolved rates are indeed measurements of $\mathrm{\Sigma_{SFR}}$ (each spaxel has an angular surface 
of 1\,arcsec$^{2}$). Obeying the Hubble flow, the distance to each galaxy is estimated and with it the correction factor for linear surface scale (kpc$^{2}$, see Fig. 
\ref{f1} and Table~\ref{tab:A1}).

\subsubsection{Global stellar mass and median age}
\label{subsec:M_est}

Total stellar masses and mean SP ages are extracted from the \textsc{starlight} output. The former are the current masses in stars whilst the latter are the mean 
light-weighted stellar ages according to \citet[][their equation 2]{Cid05}. To estimate both global M$_{*}$ and SP median age for each galaxy, we use the S/N for 
a meaningful SP fit for integrated spectra \citep[second-row criterion of Table~\ref{tab:1},][]{Cid10}. Global M$_{*}$ is the sum of 
resolved contributions whilst SP median age is the median of the age distributions of all spaxels. In our spectral sets, the restriction of above effectively omits spurious 
spectra of background galaxies but not of foreground stars so these are manually masked. Likewise the SFR, the M$_{*}$ of each star-forming region is indeed a 
measurement of $\mathrm{\Sigma}_{*}$. 

\section{Galaxy samples}
\label{sec:sam}

\subsection{The tidal perturbation parameter}
\label{subsec:f}

W-14 looked for neighbours of each CALIFA survey galaxy. In the SDSS/DR8 \citep{Aih11}, these neighbours are objects: 1) classified as galaxies, 2) within 200 kpc 
from the CALIFA survey targets, 3) with reliable values of Petrosian radii, 4) spanning sizes of at least 2 kpc, and 5) with good quality flags. 

Once the neighbours are identified, W-14 calculate what we use as criterion of segregation, the tidal perturbation parameter \citep[\textit{f},][]{Byrd86,Var04}: 
\begin{equation} \label{eq:4}
\textit{f}=\mathrm{log}\displaystyle \left(\frac{F_{prim}}{F_{sec}}\right)=3\,\mathrm{log}\displaystyle \left(\frac{R}{b}\right)+0.4\,(m_{sec}-m_{prim}).
\end{equation}

\noindent $F_{prim}$ indicates the tidal force exerted by the primary galaxy; $F_{sec}$, the internal force in the outskirts of the secondary; $m_{prim}$ and $m_{sec}$, 
their respective apparent magnitudes; \textit{R}, the secondary disk radius; and \textit{b}, the perigalactic distance of the primary. \citet{Var04} discuss that only 
for very eccentric orbits and when the primary is around the apocentre \textit{f} would fail in pointing true perturbed galaxies \citep[also see][]{Sch19}. They verify 
that errors of 20\,\% in \textit{b} and/or mass, result in errors of at most a few tenths in \textit{f}. For equal primary and secondary galaxy masses, \textit{f}$\geq$$-$2 
implies \textit{b} close enough to clearly induce global instability \citep{ByrdHow92}. On the other hand, \citet{Var04} obtain the \textit{f} distribution of the Coma 
cluster which shows no galaxies with \textit{f}$<$$-$4.5. Assuming that an aggregate as rich as a cluster is not a place for a typical non-tidally-perturbed galaxy, they 
adopt this criterion for no perturbation (so that an \textit{f}$\geq$$-$4.5 implies tidal perturbation). 

W-14 find \textit{f} and the galaxy interaction state well correlated. The latter results from separated eye-classifications of SDSS 
images (\textit{r} and \textit{i} bands) within the Collaboration. They find \textit{f}\,$=\,-$4.0 ($\sigma\,=$\,1.7) for non-interacting galaxies and \textit{f}\,$=\,-$2.9 ($\sigma\,=$\,2.0) 
for interacting ones.

\subsection{The observational strategy}
\label{subsec:strategy}

The MS was selected from the SDSS/DR7 photometric catalogue to include all galaxies with an \textit{r}-band isophotal diameter of 45''$<d<$79.2'' 
(0.005$<z<$0.03). The selection of these candidates is mainly based on visibility and to fit the PMAS/PPak field-of-view (see W-14). The PMAS \citep{Roth05}/PPak 
\citep{Kelz06,Ber10} integral field spectrograph, mounted on the Calar Alto 3.5 m telescope, was used to perform the survey observations. Mostly all these were 
selected from the MS. A three dithering scheme was adopted to fill the gaps among PPak fibers. With this, the vignetting trouble and spatial resolution are 
respectively reduced and improved \citep{San12a}. Two different but complementary set ups were used to perform all observations. The one of medium resolution 
is used here ($\lambda/\Delta\lambda\,\sim$850 at $\sim$5\,000\,\AA{}, FWHM $\sim$6\,\AA{}). Its main purpose is to study the SPs and the properties of the 
ionized gas by including as much emission line species as possible (\textit{i.e.}, the widest wavelength range, 3745-7300\,\AA{}).

The improved spectrophotometric calibration and registering of the images are finally remarked. The scaling of the absolute flux levels of the datacubes to 
SDSS/DR7 broad-band photometry is better than a 15\,\% \citep[DR1,][]{Hus13}. Later, it improves to $\sim$8\,\% due to a new registering procedure \citep[DR2, 
DR3,][]{Gar15,San16a}. Predicted SDSS fluxes for the PPak fibers are compared with those of the spaxel spectra themselves at each pointing position. The 
photometric scale factor at the best matching position is used to rescale the absolute photometry of each particular pointing of the spectra. The reliability 
of the nebular fluxes is reinforced\footnote{Aperture size corrections are even needless: $\sim$97\,\% of CALIFA survey galaxies are covered out to at least 
2 $\times$ the SDSS Petrosian half light radius as computed from the growth curve photometry \citep[see][]{San16a}.}. However, for irregular cases such as 
edge-on galaxies, the registering fails so the previous calibration is re-used. 

\subsection{The selection of the samples}
\label{subsec:gaxsamp}

The CALIFA survey consists of 667 objects\footnote{https://bit.ly/2IeelCX}. From them, 542 are a subset of the MS. The fraction of 529 out of 542 was observed in 
the widest wavelength range. From it, 454 objects have \textit{f} estimations (W-14). Under the criterion of \citet{Var04}, \textit{i.e.} \textit{f}$<$$-$4.5 for 
no perturbation, 101 are non-tidally perturbed and 353 are tidally perturbed.

\begin{figure}
   \mbox{\includegraphics[width=.4948\columnwidth]{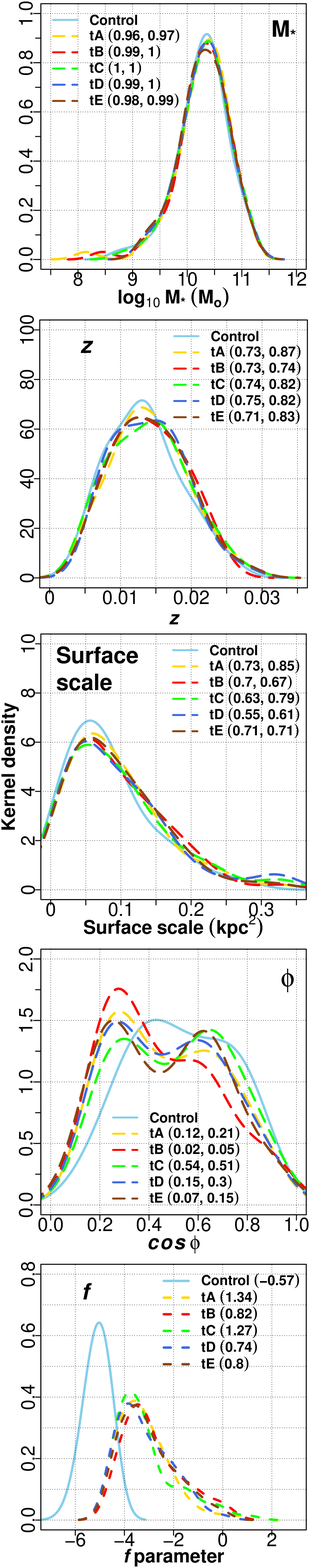}}
   \mbox{\includegraphics[width=.4948\columnwidth]{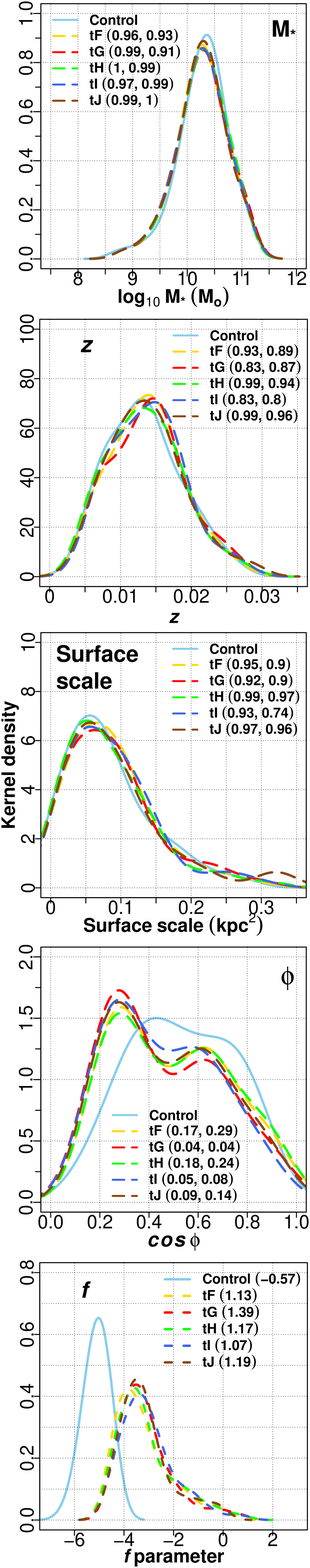}}
\caption{\scriptsize{Fundamental properties: Kernel density estimations for the distributions of fundamental M$_{*}$ and \textit{z} (top). Distributions of 
surface scale, inclination angle ($\phi$) and \textit{f} parameter are also included (middle to bottom). The perturbed samples, trials A to J, are shorten 
as tA to tJ. For M$_{*}$ to $cos\,\phi$, results from the Anderson-Darling (AD) and permutation (equal densities) tests are shown for each perturbed sample 
against the control one. The skewness ($S$) as a normality test, is shown for the \textit{f} parameter. Decisions are fairly normal ($-0.5\leq\,S$ 
$\leq0.5$), moderately skewed ($-1\leq\,S<\,-0.5$, $0.5\,<S\,\leq1$) and highly skewed ($S<\,-1$, $S>\,1$).}}
   \label{f1} 
\end{figure}

We obtain the resolved BPT diagram, with a median star-forming region fraction of 0.85, for 62 out of the 101 non-tidally-perturbed objects (61\,\%). Such 62 ones 
are then ELGs (see Section \ref{subsec:s-f_rsub}) and 41/62 of them are Late Type Spirals (LTSs, see Table \ref{tab:A1}). The \textit{f} parameter  
cumulative distribution function of the 62 objects is also nearly normal (see Fig.~\ref{f1}, bottom). This effective set of non-tidally perturbed objects is called 
hereafter the \emph{control} sample. Similarly, we obtain the resolved BPT diagram, with 0.81 as the median fraction of star-forming regions, for 231 out of the 
353 tidally-perturbed objects (65\,\%). These are ELGs as well. To perform fair comparisons of SP properties, we construct from these perturbed ELGs, ten subsets 
with 62 objects each that mimic as close as possible five fundamental properties of the control sample. These properties are M$_{*}$, redshift 
(\textit{z}), morphological group, galaxy colour and the dominant excitation source (see Table~\ref{tab:A1}). The reasons behind these fundamentals are the following.
\begin{enumerate}
 \item M$_{*}$ to control differences in SF records. 
 \item \textit{z} to minimize contrasts in spatial or surface scales and number of star-forming regions.
 \item Morphological group since morphology contributes (at least) to stabilize SF (\textit{e.g.} \citealt{Mar09}, \citealt{Mar13}, \citealt{Gon15}, \citealt{Can19}, 
 \citealt{Men19}, \citealt{Koy19}).
 \item Galaxy colour for homogeneous photometry.
 \item The dominant source of gas excitation to balance the influence of a plausible massive black hole \citep[AGN feedback, \textit{e.g.}][]{Schaw07,Fab12,Cic14}.
\end{enumerate}

From the 231 perturbed ELGs, the ten samples, trials A to J (62 objects each), are built in a two-step procedure. Trials A to E are obtained first sort ascending 
the control sample by M$_{*}$\footnote{M$_{*}$ is preferred, for instance, over \textit{z}, since it is easier to compare a quantity which has no need of more than 
two tenths to be significant (\textit{z} is often expressed with more than two tenths).}. The most approximated case in the above five properties to each single 
control object is selected. The trials are filled-in simultaneously in order to avoid, when possible, common cases within them. Secondly, trials F to J are obtained, 
this time sort descending the M$_{*}$ of the control sample. Trials A to E use a total of 139/231 (60\,\%) perturbed galaxies with 13 common cases within them five. 
Likewise, trials F to J use 133/231 (58\,\%) perturbed galaxies with 17 common cases within them all. Notice a trend of increasing common cases as the fraction of 
usage decreases\footnote{A third set of five trials (not shown) with a usage of 150/231 (65\,\%) has 9 common cases. However, beside the control sample, it has the 
less alike \textit{z} distributions, the more unlike frequencies of objects according to the source of gas excitation, and so on.}. In sum, the ten trials have a 
usage of 162/231 (70\,\%, see Table~\ref{tab:A1}).
               
\begin{figure}
   \mbox{\includegraphics[width=.4948\columnwidth]{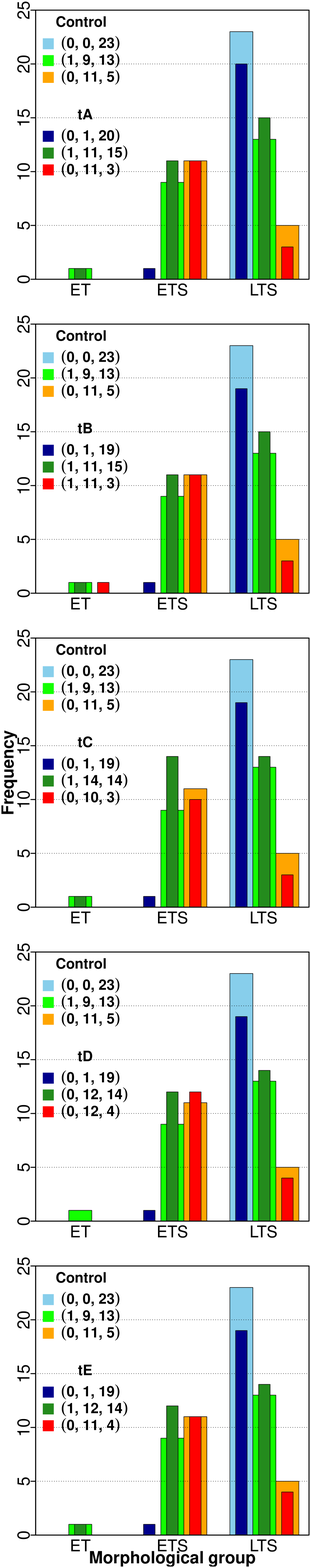}}
   \mbox{\includegraphics[width=.4948\columnwidth]{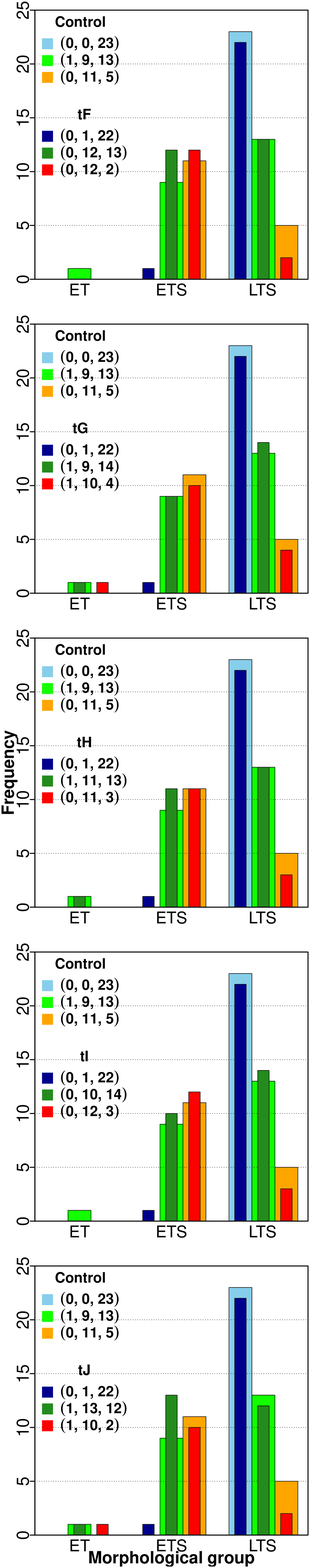}}
\caption{\scriptsize{Fundamental properties: the morphological group-galaxy colour relation between control (Control) and perturbed samples (tA to tJ). Colour 
frequencies in parenthesis are ordered by morphological group (\textit{i.e.} early type, ET; early type spiral, ETS; late type spiral, LTS).}}
   \label{f2} 
\end{figure}

\begin{figure}
   \mbox{\includegraphics[width=.4948\columnwidth]{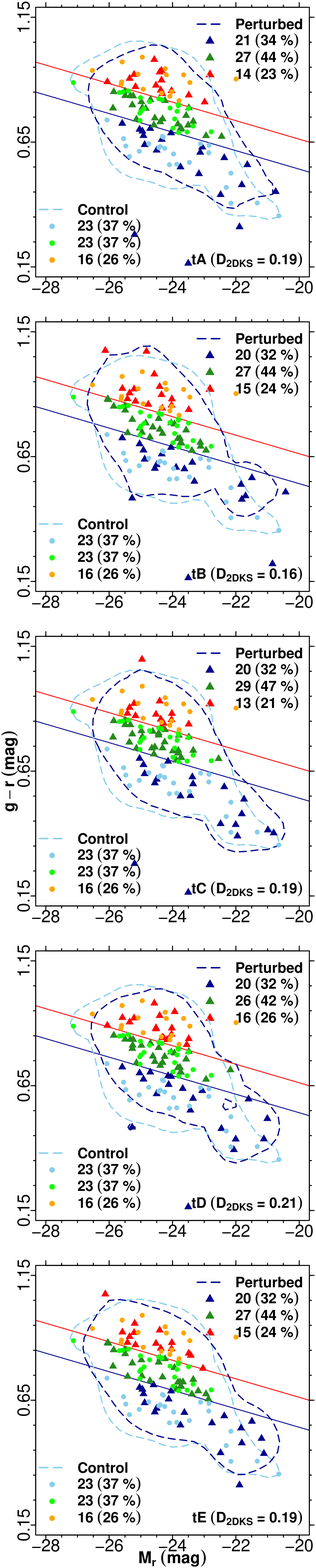}}
   \mbox{\includegraphics[width=.4948\columnwidth]{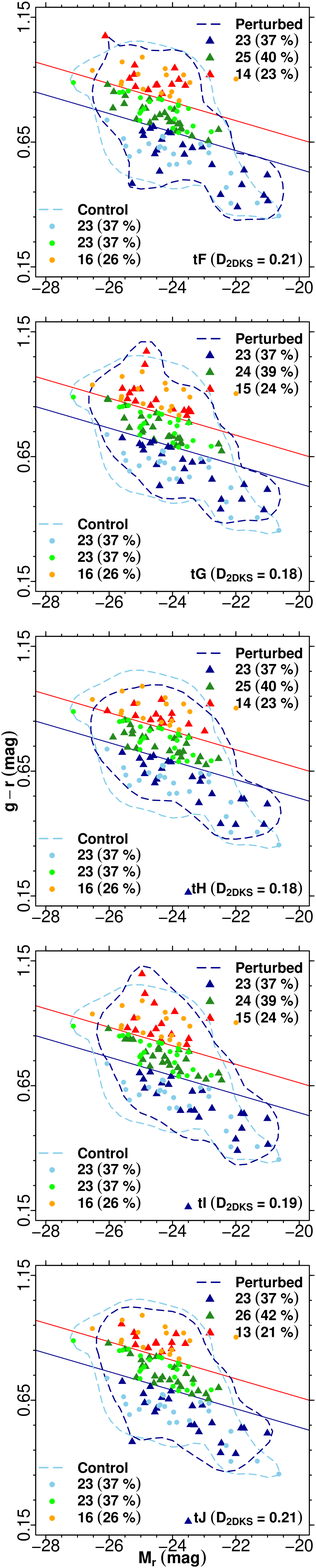}}
\caption{\scriptsize{Fundamental properties: CMDs showing the distributions for control (Control) and perturbed samples (tA to tJ). The colour solid lines 
are explained in Section~\ref{subsec:gal-col}. The number of colour galaxies with their respective percentages, as well as 0.1 Kernel density contours 
(dashed lines) are indicated for each sample. From the 2D K-S/Peacock two-sample test, each bottom-right corner gives the maximum absolute difference 
(D$_{\mathrm{2DKS}}$) between each empirical cumulative probability distribution.}}
   \label{f3} 
\end{figure}

Figure~\ref{f1} (top) shows the contrasts in distributions for fundamental M$_{*}$ and \textit{z}. For the former, the ten perturbed samples tightly follow the 
distribution of the control one whilst results of the statistical tests are nearly one. For \textit{z}, the dashed-line distributions closely follow the control one 
with statistical results of at least 0.7. The rest properties, surface scale, \textit{cos} $\phi$ and \textit{f} parameter are later treated in Section \ref{sec:val}. 
Figure \ref{f2} illustrates the morphological group-galaxy colour relation. Notice that all panels show an increment of blue objects and a decrement of red 
ones as the morphological group becomes later. Frequencies of all perturbed samples by group and colour little differ from those of the control one. Figure~\ref{f3} 
plots the CMD for each pair of comparisons. Similarly, frequencies and percentages of colour perturbed galaxies are well balanced with respect to those of 
control objects. Moreover, most of the area enclosed by the density contours overlap. At the statistical level, differences between each pair of probability distributions 
(D$_{\mathrm{2DKS}}$, bottom-right) support the null hypothesis of a common parent distribution. Finally, the frequencies related to the dominant source of gas excitation 
are listed. The control sample consists of 52/10 SFG/AGN-like type objects. In the same way, contents of samples tA to tJ are 50/12, 51/11, 51/11, 52/10, 51/11, 49/13, 
50/12, 50/12, 50/12, 49/13 SFG/AGN-like types respectively.

\subsection{Sample comparison}
\label{sec:val}

Contrasts in spatial scales of the single star-forming regions and in galaxy inclinations with respect to the plane of the sky can be both sources 
of observational bias. Density estimations for the spatial or surface scale, \textit{cos}\,$\phi$ and \textit{f} parameter distributions are shown in Fig.~\ref{f1} 
(see also Table \ref{tab:A1}). For surface scales, the dashed-line distributions resemble the control one with statistical results of at least 0.6. These scales 
directly depend on \textit{z} so the density estimations and statistical results of the scales will always be as good as those of \textit{z}. In contrast, the 
\textit{cos}\,$\phi$ distributions are the odd case. The density estimations evidently differ: the perturbed samples are biased towards larger inclinations. We 
therefore relocate the coordinates of each spaxel by deprojecting them on those galaxies that show a disk component. The number of non-deprojected cases is larger 
for the perturbed samples (tA to tJ) consequence of their higher bias. Non-deprojected cases, \textit{i.e.} low \textit{cos}\,$\phi$ values for trials tA to tJ are 13, 17, 8, 
11, 15, 12, 16, 13, 13, and 17 respectively. Likewise, they are 6 for the control sample. Regarding the \textit{f} parameter distributions, skewness is used as 
a measure of normality. The control sample distribution, close to be fairly normal, is moderately (negatively) skewed. On the contrary, the distributions of the perturbed 
samples are positively skewed, most of them highly (outliers).

Having shown alike surface scales for resolved regions, we now explore contrasts in frequencies of star-forming regions among samples and per 
single galaxy. Table~\ref{tab:2} lists these frequencies by splitting the control and perturbed samples according to the source of gas excitation. No known phenomenon 
prevents most of the available gas to turn into stars in SFG Blue and SFG LTS objects so both are the ones shown apart only for the sake of briefness. SFG Green and 
SFG Red types are included in the SFG subsample and are shown apart from Section~\ref{sec:res} on. Even fairer comparisons will be conducted adopting this subsampling 
(not done in AGN-like galaxies due to the size limitation). Notice, from Table~\ref{tab:2} (top), well balanced galaxy frequencies. In the case of the totals of 
star-forming regions (Table~\ref{tab:2}, middle), disproportions are the lack of SFG Blue early type spirals (ETSs) in control galaxies and the AGN-like case of perturbed 
galaxies almost doubling those frequencies of control ones (by a factor of $\sim$1.9). This last contrast is not expected since differences in galaxy frequencies are 
not as great as double. Even the respective \textit{z} distributions exhibit no marked biases. Moreover, excluding the AGN-like subsample (with factors of $\sim$1.3 
and $\sim$1.6), median and mean frequencies per single galaxy (Table~\ref{tab:2}, bottom) differ not much (largest factors are of $\sim$1.1). Anderson-Darling (AD) tests for the distributions 
of numbers of star-forming regions support this. Likelihoods are 0.59, 0.58, 0.75 and 0.48 for SFG Blue, SFG LTS, SFG and AGN-like respectively. In sum, numbers of 
star-forming regions in AGN-like galaxies are the most dissimilar ones. 

\begin{table}
   \setlength{\tabcolsep}{0.75\tabcolsep}
 \begin{minipage}{\columnwidth}
 \caption{\scriptsize{Frequency summaries of galaxies (top) and star-forming regions (middle) as defined in Section~\ref{subsec:s-f_rsub}. Star-forming region statistics 
 per single galaxy (bottom).}
 \label{tab:2}}
 \centering
 \begin{scriptsize}
 \begin{tabular}{l@{\hspace{1.5\tabcolsep}}c@{\hspace{1.5\tabcolsep}}c@{\hspace{1.5\tabcolsep}}c@{\hspace{1.5\tabcolsep}}c@{\hspace{1.5\tabcolsep}}c@{\hspace{1.5\tabcolsep}}c@{\hspace{1.5\tabcolsep}}c@{\hspace{1.5\tabcolsep}}c@{\hspace{1.5\tabcolsep}}c}
 \hline
          &                                \multicolumn{4}{c}{Control}                                &&\multicolumn{4}{c}{Perturbed\footnote{\label{Missed}\scriptsize{Median values from gathering trials A to J (tA to tJ).}}}\\[1ex]
          &SFG                   &SFG                   &                      &AGN                   &&SFG                   &SFG                   &                      &AGN                   \\
          &Blue                  &LTS                   &SFG                   &-like                 &&Blue                  &LTS                   &SFG                   &-like                 \\
\cline{2-5}\cline{7-10}                                                                                                                                                                            \\
ETSs      &$\ldots$              &$\ldots$              &12                    &\phantom{0}9          &&\phantom{0}1          &$\ldots$              &13                    &11                    \\
LTSs      &23                    &40                    &40                    &\phantom{0}1          &&20                    &37                    &37                    &1                     \\
Total     &23                    &40                    &52                    &10                    &&21                    &37                    &50                    &12                    \\[1ex]
\cline{2-5}\cline{7-10}                                                                                                                                                                            \\                                                                                                                       
ETSs      &$\ldots$              &$\ldots$              &\phantom{0}4\,396     &1\,024                &&\phantom{0}1\,462     &$\ldots$              &\phantom{0}5\,818     &1\,900                \\
LTSs      &21\,223               &31\,001               &31\,001               &\phantom{0}\,218      &&20\,803               &31\,907               &31\,907               &\phantom{0}\,412      \\
Total     &21\,223               &31\,001               &35\,397               &1\,242                &&22\,265               &31\,907               &37\,725               &2\,312                \\[1ex]
\cline{2-5}\cline{7-10}                                                                                                                                                                            \\
median    &925                   &746                   &628                   &\phantom{0}98         &&\phantom{0}\,894      &713                   &628                   &127                   \\
mean      &923                   &775                   &681                   &124                   &&1\,060                &862                   &754                   &193                   \\
$\sigma$  &442                   &428                   &435                   &\phantom{0}93         &&\phantom{0}\,664      &599                   &586                   &157                   \\
\hline
 \end{tabular}
 \end{scriptsize}
 \end{minipage}
 \end{table}

\subsection{H$\alpha$ flux percentage radii}
\label{subsec:annuli&profiles}

\begin{figure}\centering
   \mbox{\includegraphics[width=.494438\columnwidth]{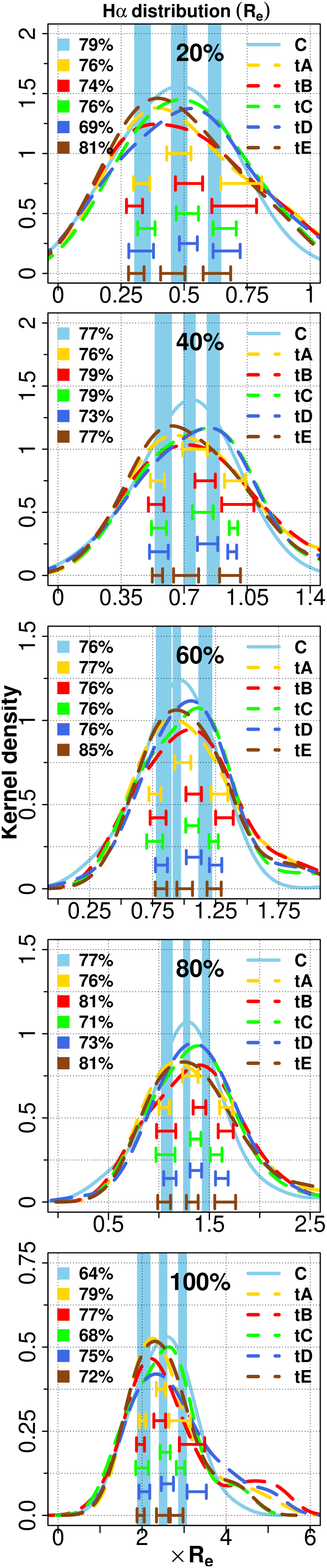}}
   \mbox{\includegraphics[width=.494438\columnwidth]{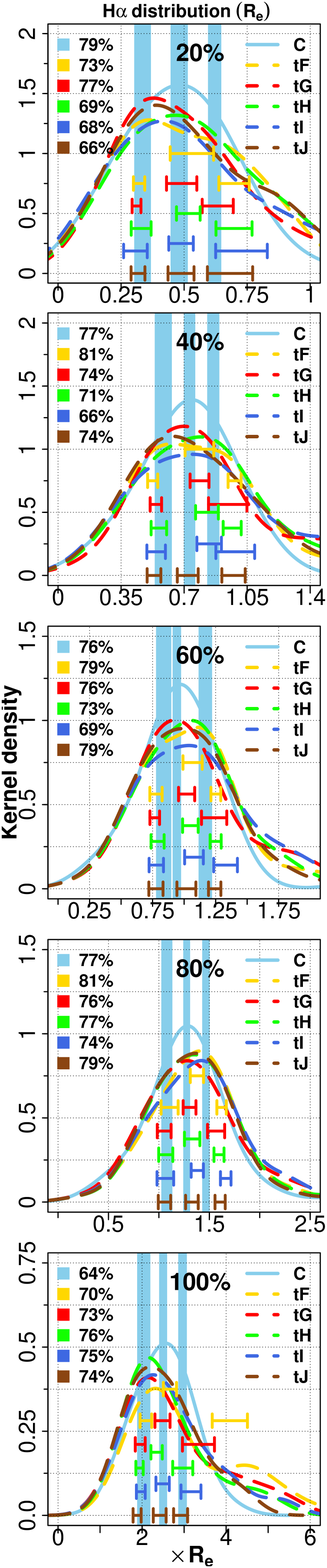}}
\caption{\scriptsize{Kernel density estimations for our H$\alpha$-flux-percentage radius/R$_{e}$ fractions, \textit{i.e.}, the H$\alpha$ emission line distribution 
as a function of R$_{e}$ (20 to 100\,\%, from top bottom). Control (C) and perturbed samples, trials A to E and F to J (tA to tE, left and tF to tJ, right). At each left 
are the percentages of fractions (one fraction per galaxy) within $\pm$\,1\,$\sigma$ width centred at each distribution median. Light-blue columns are the Monte Carlo 
standard deviations (MCsds, \textit{i.e.} $\pm$\,1\,$\sigma$ width from Monte Carlo simulations) corresponding to and centred at the 1st, 2nd (median) and 3rd quartiles 
(Qs) of the control distribution. Likewise, coloured ranges represent the same for each perturbed sample distribution.}}
   \label{f4} 
\end{figure}

All annular profiles compile single-galaxy data and depict each sample or subsample SP properties by concentric annuli. These are 
defined by the boundaries of deprojected radii which encircle percentages of the all-excitation H$\alpha$ flux (20, 40, 60, 80 and 100\,\%). From each galaxy set 
of spectra solved by \textsc{starlight}, the total H$\alpha$ flux is computed by ignoring the excitation source and flux outliers. The flux percentages are then 
computed and so the encircling radii. Since such those fluxes are meant for sectioning only, they are not extinction-corrected. Besides, their respective spectra 
sometimes show no H$\beta$ line emission detections (poor line S/N ratios). See Appendix~\ref{sec:app2} for an additional but important note on these annular profiles.

As noticed already, the H$\alpha$ flux, related to the data per se, is used to depict the radial extension instead of photometric extents such as the effective 
radius (R$_{e}$). To prove the reliability of the encircling-H$\alpha$-flux radii, Fig.~\ref{f4} plots Kernel density estimations of the H$\alpha$ emission line 
distribution of our sampled galaxies as a function of R$_{e}$ (half-light radius of the \textit{r} band in an elliptical aperture, see W-14). Table~\ref{tab:3} 
lists medians, 3rd quartiles (Qs) and standard deviations. Statistical test results between samples for the sets of these radial fractions are also included. At 
each percentage radius/R$_{e}$ fraction in Fig.~\ref{f4} (panels 20 to 100\,\%), numbers at left indicate the percentages of fractions (one 
fraction per galaxy) close to the median of each respective set, \textit{i.e.}, within $\pm$\,1\,$\sigma$ width range centred in the median. In general, a bit 
more than a 70\,\% of the fractions are near the median in each distribution. Moreover, medians of fractions in Table~\ref{tab:3} are in good agreement with 
the standard definition of R$_{e}$. For instance, medians for the mid-radius (60\,\%) are the ones closest to unity (see also Table~\ref{tab:A1}). These H$\alpha$ 
flux percentage radii are then credible distance normalizers and may be used as indicators of galaxy extensions. 

From Monte Carlo simulations, standard deviations (MCsds) for medians, 1st and 3rd Qs are obtained for the distributions of percentage radius/R$_{e}$ fractions. 
They are drawn as columns for the control sample, and as ranges for trials tA to tJ (see Fig.~\ref{f4}). If contrasting the ranges with respect to the columns, 
the former are biased towards higher fractions, specifically, medians and 3rd Qs. Also see that the curves are much different from the 1st Q on (right skewed). 
If comparing medians and 3rd Qs by computing fractions from Table~\ref{tab:3}, we find no significant difference for medians: 31/50 fractions advantage perturbed 
samples by $<$10\,\%. Third Qs are slightly different: 27/50 fractions advantage perturbed samples by $>$10\,\%. Differences in standard deviations are the most 
significant: 33/50 advantage perturbed galaxies by $>$20\,\%. We look at last at the statistical tests (AD-P) of Table~\ref{tab:3}. If arranging the likelihoods in 
the $l>$0.5, 0.25$<l\leq$0.5 and $l\leq$0.25 groups, frequencies for AD-permutation tests are respectively 6-4, 15-17 and 29-29. A likelihood for similitude of 
$\leq$0.25 describes the most both sample distribution functions. In sum, it may be said that the H$\alpha$ line emission appears to be a little more dispersed 
for perturbed galaxies.

\section{Results}
\label{sec:res}

\begin{table}
 \begin{minipage}{\columnwidth}
 \caption{\scriptsize{Medians, 3rd quartiles (Qs) and standard deviations (median-3rdQ-$\sigma$) of the percentage radius/R$_{e}$ fractions (\textit{e.g.} 
 $\times$\,R$_{e}$). AD and permutation test results (AD-P) for each radial set of fractions between control and each perturbed sample (tA to tJ) are also 
 listed.}
 \label{tab:3}}
  \centering
 \begin{scriptsize}
 \begin{tabular}{c@{\hspace{0.55\tabcolsep}}c@{\hspace{0.55\tabcolsep}}c@{\hspace{0.55\tabcolsep}}c@{\hspace{0.55\tabcolsep}}c}
 \hline
  \multicolumn{5}{c}{H$\alpha$ flux percentage radii} \\
20\%      &40\%      &60\%      &80\%      &100\%     \\
\hline                                                
                          \multicolumn{5}{c}{Control}                          \\
0.48-0.62-0.27 &0.73-0.86-0.30 &0.94-1.17-0.33 &1.28-1.47-0.40 &2.50-2.96-0.63 \\
                            \multicolumn{5}{c}{tA}                             \\
0.48-0.73-0.33 &0.76-0.98-0.38 &0.99-1.28-0.40 &1.31-1.67-0.48 &2.45-2.89-0.89 \\
0.27-0.26      &0.22-0.18      &0.20-0.24      &0.16-0.12      &0.41-0.18      \\
                            \multicolumn{5}{c}{tB}                             \\
0.52-0.70-0.34 &0.82-1.00-0.44 &1.08-1.32-0.49 &1.40-1.66-0.60 &2.42-3.19-1.09 \\
0.24-0.23      &0.11-0.11      &0.06-0.10      &0.06-0.09      &0.21-0.06      \\
                            \multicolumn{5}{c}{tC}                             \\
0.51-0.66-0.29 &0.80-0.97-0.36 &1.06-1.23-0.38 &1.36-1.57-0.43 &2.55-2.91-0.77 \\
0.54-0.80      &0.21-0.36      &0.17-0.28      &0.23-0.37      &0.81-0.76      \\
                            \multicolumn{5}{c}{tD}                             \\
0.52-0.67-0.27 &0.83-0.97-0.33 &1.08-1.25-0.35 &1.37-1.62-0.38 &2.60-3.30-0.98 \\
0.46-0.49      &0.24-0.28      &0.24-0.44      &0.26-0.34      &0.18-0.16      \\
                            \multicolumn{5}{c}{tE}                             \\
0.45-0.63-0.31 &0.71-0.95-0.38 &1.00-1.24-0.42 &1.33-1.66-0.56 &2.49-2.82-0.75 \\
0.81-0.61      &0.36-0.28      &0.19-0.51      &0.12-0.23      &0.75-0.39      \\
                            \multicolumn{5}{c}{tF}                             \\
0.53-0.70-0.31 &0.79-0.98-0.42 &1.07-1.25-0.47 &1.38-1.62-0.62 &2.66-4.09-1.31 \\
0.28-0.25      &0.17-0.21      &0.12-0.19      &0.09-0.30      &0.02-0.01      \\
                            \multicolumn{5}{c}{tG}                             \\
0.49-0.63-0.29 &0.79-0.94-0.37 &1.02-1.24-0.45 &1.30-1.57-0.52 &2.49-3.34-1.14 \\
0.51-0.36      &0.21-0.16      &0.22-0.24      &0.44-0.38      &0.10-0.01      \\
                            \multicolumn{5}{c}{tH}                             \\
0.52-0.70-0.28 &0.83-0.97-0.34 &1.05-1.25-0.37 &1.33-1.60-0.46 &2.35-2.97-0.96 \\
0.32-0.30      &0.22-0.21      &0.20-0.24      &0.34-0.39      &0.27-0.04      \\
                            \multicolumn{5}{c}{tI}                             \\
0.49-0.73-0.34 &0.84-0.98-0.41 &1.08-1.33-0.46 &1.38-1.66-0.51 &2.49-3.16-1.00 \\
0.23-0.11      &0.08-0.02      &0.05-0.03      &0.08-0.13      &0.29-0.15      \\
                            \multicolumn{5}{c}{tJ}                             \\
0.49-0.68-0.27 &0.72-0.97-0.36 &1.02-1.24-0.42 &1.33-1.61-0.47 &2.42-2.92-0.82 \\
0.43-0.24      &0.29-0.11      &0.28-0.32      &0.33-0.46      &0.54-0.25      \\
\hline\\
 \end{tabular}
 \end{scriptsize}
 \end{minipage}
 \end{table}

\subsection{Resolved SFMSs}
\label{subsec:mssf_1}

\begin{table}
   \setlength{\tabcolsep}{0.5\tabcolsep}
 \begin{minipage}{\columnwidth}
\caption{\scriptsize{Frequency summary from the annular comparisons between samples for all subsamples (see Table~\ref{tab:C1}). Numbers in parenthesis are totals 
of star-forming regions and annular means. Columns titled as ``Fig.~\ref{f5}'' list the frequencies from the comparisons in Fig.~\ref{f5}. Acronyms (first column) 
mean frequencies of Different Slopes (DS), of Perturbed samples having the Flattest slope (PF) and frequencies of Different Distribution functions (DD).}
 \label{tab:4}}
  \centering
 \begin{scriptsize}
 \begin{tabular}{l@{\hspace{0.75\tabcolsep}}c@{\hspace{0.3\tabcolsep}}c@{\hspace{0.3\tabcolsep}}c@{\hspace{0.3\tabcolsep}}c@{\hspace{0.3\tabcolsep}}c@{\hspace{0.3\tabcolsep}}c@{\hspace{0.3\tabcolsep}}c@{\hspace{0.75\tabcolsep}}c@{\hspace{0.3\tabcolsep}}c@{\hspace{0.3\tabcolsep}}c@{\hspace{0.3\tabcolsep}}c@{\hspace{0.3\tabcolsep}}c@{\hspace{0.3\tabcolsep}}c}
 \hline
   &                \multicolumn{5}{c}{Annuli (H$\alpha$ flux percentages)}                &Fig.     &&                \multicolumn{5}{c}{Annuli (H$\alpha$ flux percentages)}                 &Fig.     \\
   &20\%            &40\%             &60\%             &80\%             &100\%           &\ref{f5} &&20\%            &40\%             &60\%             &80\%             &100\%            &\ref{f5} \\
\cline{2-7}\cline{9-14}                                                                                                                                                                                  \\
   &                       \multicolumn{5}{c}{AGN-like (36\,552, 7\,310)}                  &         &&                     \multicolumn{5}{c}{SFG Red (68\,654, 13\,731)}                     &         \\[1ex]
DS &10/10           &\phantom{0}7/10  &\phantom{0}3/10  &\phantom{0}7/10  &10/10           &4/5      &&\phantom{0}8/10 &\phantom{0}6/10  &\phantom{0}4/10  &\phantom{0}7/10  &10/10            &4/5      \\
PF &\phantom{0}0/10 &\phantom{0}0/7   &\phantom{0}3/3   &\phantom{0}1/7   &10/10           &1/4      &&\phantom{0}3/8  &\phantom{0}4/6   &\phantom{0}0/4   &\phantom{0}0/7   &\phantom{0}3/10  &1/4      \\
DD &\phantom{0}7/10 &9/10             &\phantom{0}9/10  &9/10             &\phantom{0}10/10&5/5      &&10/10           &9/10             &\phantom{0}9/10  &\phantom{0}8/10  &\phantom{0}10/10 &5/5      \\[1ex]
   &                      \multicolumn{5}{c}{SFG ETS (122\,322, 24\,464)}                  &         &&                     \multicolumn{5}{c}{SFG Green (208\,180, 41\,636)}                  &         \\[1ex]
DS &\phantom{0}8/10 &10/10            &9/10             &\phantom{0}8/10  &\phantom{0}10/10&5/5      &&\phantom{0}8/10 &\phantom{0}6/10  &\phantom{0}3/10  &\phantom{0}4/10  &\phantom{0}8/10  &4/5      \\
PF &\phantom{0}1/8  &\phantom{0}0/10  &\phantom{0}0/9   &\phantom{0}0/8   &\phantom{0}0/10 &0/5      &&\phantom{0}0/8  &\phantom{0}2/6   &\phantom{0}3/3   &\phantom{0}4/4   &\phantom{0}8/8   &3/4      \\
DD &10/10           &10/10            &\phantom{0}9/10  &10/10            &\phantom{0}10/10&5/5      &&10/10           &\phantom{0}9/10  &\phantom{0}9/10  &10/10            &\phantom{0}8/10  &5/5      \\[1ex]
   &                     \multicolumn{5}{c}{SFG Blue (376\,698, 75\,340)}                  &         &&                    \multicolumn{5}{c}{SFG LTS (531\,210, 106\,242)}                    &         \\[1ex]
DS &\phantom{0}9/10 &10/10            &10/10            &\phantom{0}2/10  &\phantom{0}3/10 &3/5      &&10/10           &10/10            &10/10            &\phantom{0}6/10  &\phantom{0}5/10  &5/5      \\
PF &\phantom{0}9/9  &10/10            &10/10            &\phantom{0}1/2   &\phantom{0}2/3  &3/3      &&10/10           &10/10            &10/10            &\phantom{0}6/6   &\phantom{0}4/5   &5/5      \\
DD &9/10            &10/10            &10/10            &\phantom{0}9/10  &\phantom{0}9/10 &5/5      &&10/10           &10/10            &10/10            &\phantom{0}6/10  &10/10            &5/5      \\[1ex]
   &                       \multicolumn{5}{c}{SFG (653\,532, 130\,706)}                    &         &&                        \multicolumn{5}{c}{All (690\,084, 138\,017)}                    &         \\[1ex]
DS &\phantom{0}9/10 &10/10            &\phantom{0}10/10 &\phantom{0}5/10  &\phantom{0}5/10 &5/5      &&\phantom{0}9/10 &10/10            &10/10            &\phantom{0}7/10  &\phantom{0}8/10  &5/5      \\
PF &\phantom{0}9/9  &10/10            &\phantom{0}10/10 &\phantom{0}5/5   &\phantom{0}4/5  &5/5      &&\phantom{0}9/9  &10/10            &10/10            &\phantom{0}7/7   &\phantom{0}8/8   &5/5      \\
DD &10/10           &10/10            &10/10            &\phantom{0}8/10  &10/10           &5/5      &&10/10           &10/10            &10/10            &\phantom{0}9/10  &10/10            &5/5      \\[1ex]
\hline\\
 \end{tabular}
 \end{scriptsize}
 \end{minipage}
 \end{table}

\begin{figure*}\centering
   \mbox{\includegraphics[width=.51665\columnwidth]{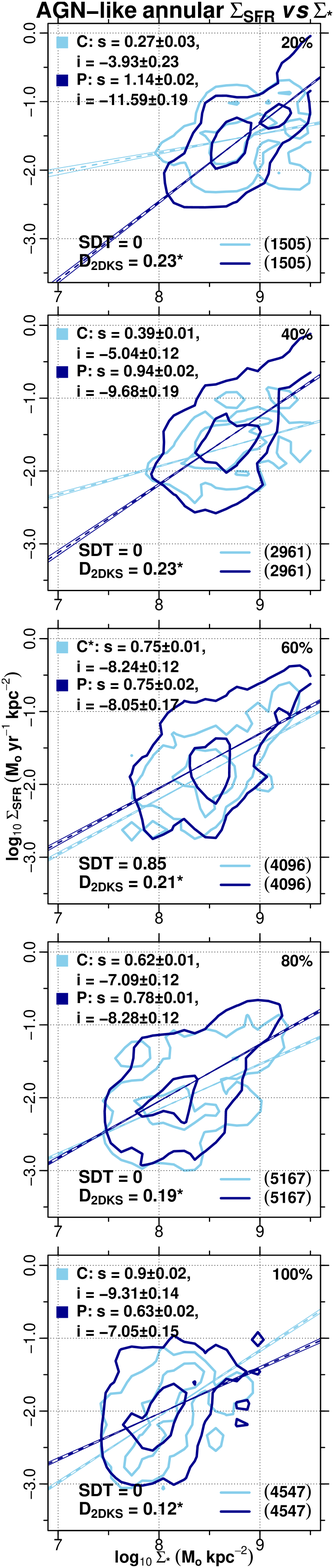}}
   \mbox{\includegraphics[width=.51665\columnwidth]{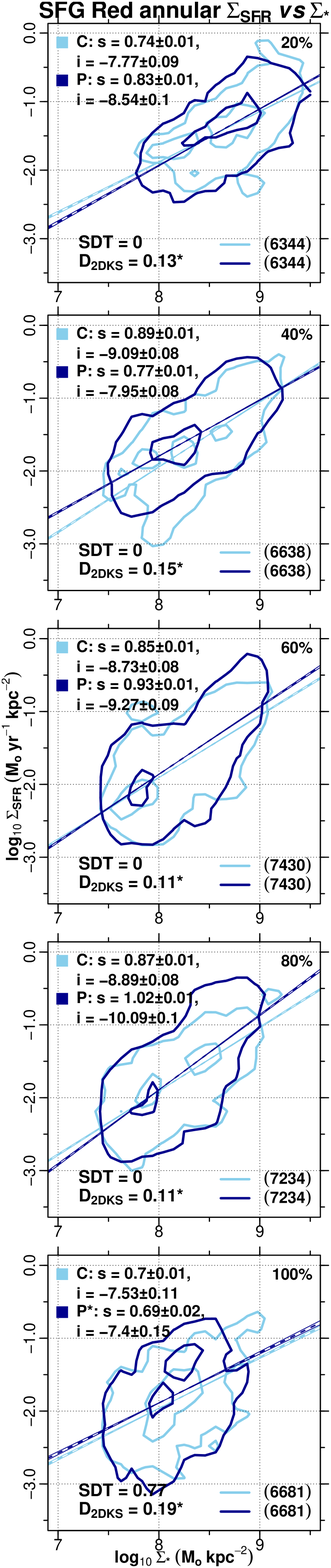}}
   \mbox{\includegraphics[width=.51665\columnwidth]{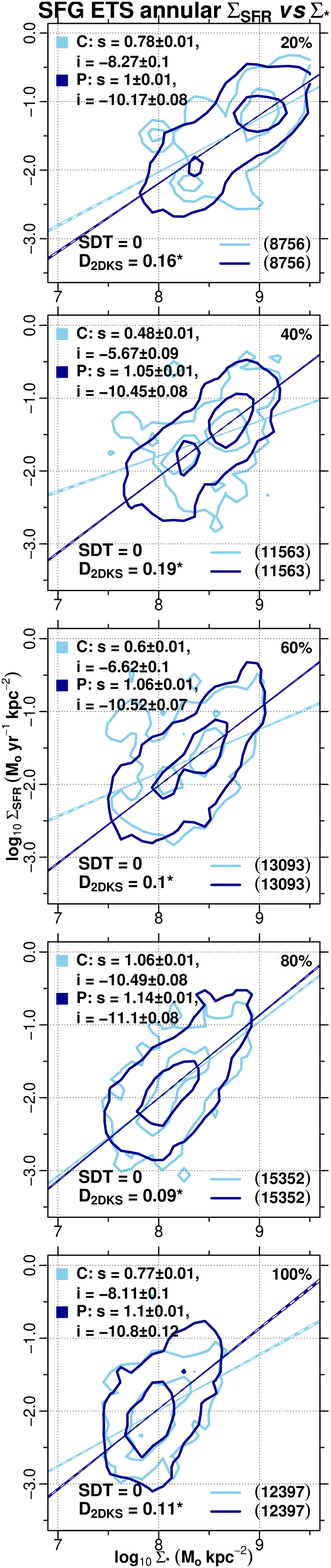}}
   \mbox{\includegraphics[width=.51665\columnwidth]{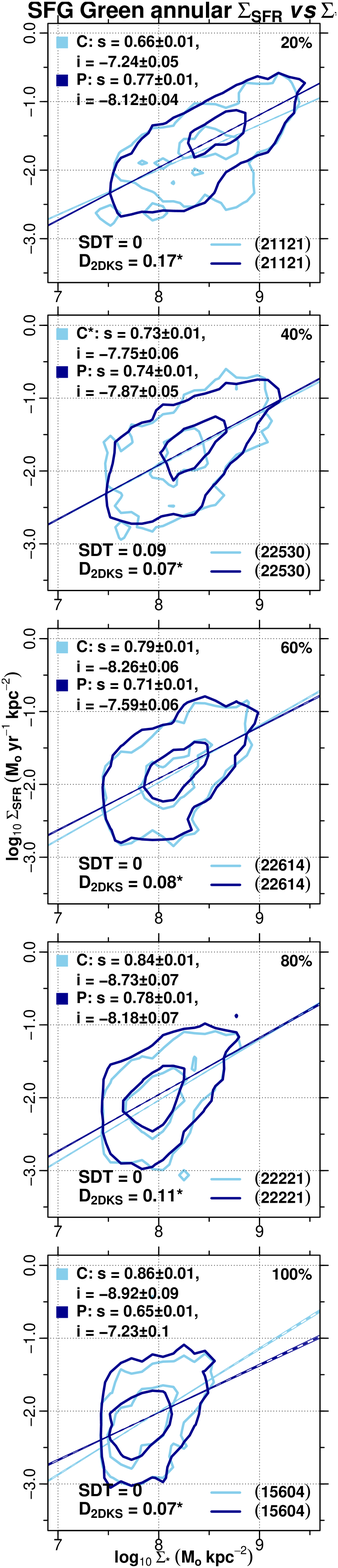}}
\caption{\scriptsize{Annular-resolved SFMSs. The perturbed star-forming regions (dark blue) are paired with their control ones (light blue) closest in $\mathrm{\Sigma}_{*}$ 
(\textit{i.e.}, paired by the minimum $\mathrm{\Sigma}_{*}$ difference). All star-forming regions in each comparison set (\textit{i.e.} trials tA to tJ 
each one with its respective control sample regions, see Appendix~\ref{sec:app3}) are considered in order to get average linear model coefficients. From left 
to right: Kernel density estimations (0.1 and 0.9 contour densities from outside-in) of the star-forming regions inhabiting AGN-like, SFG Red, SFG ETS and SFG 
Green galaxies. Control (C) and perturbed (P) linear regression coefficients: slope (\textit{s}) and intercept (\textit{i}) (models for all annular sequences 
are statistically significant). Though not perceptible in all annuli, the error intervals are drawn for all fits. Asterisks indicate, from significance difference 
tests (SDTs), the sample (C or P) with \textit{s} closer to that one if combining both-sample regions (Section~\ref{subsec:representatives}). Likewise, SDTs are 
shown for each pair of sample slopes. 2D K-S/Peacock two-sample test differences \citep[$\mathrm{D_{2DKS}}$,][]{Yua17} are finally included. Asterisks indicate 
that both distribution functions come from a different parent distribution.}}
   \label{f5} 
\end{figure*}

\begin{figure*}\centering
\ContinuedFloat   
   \mbox{\includegraphics[width=.51665\columnwidth]{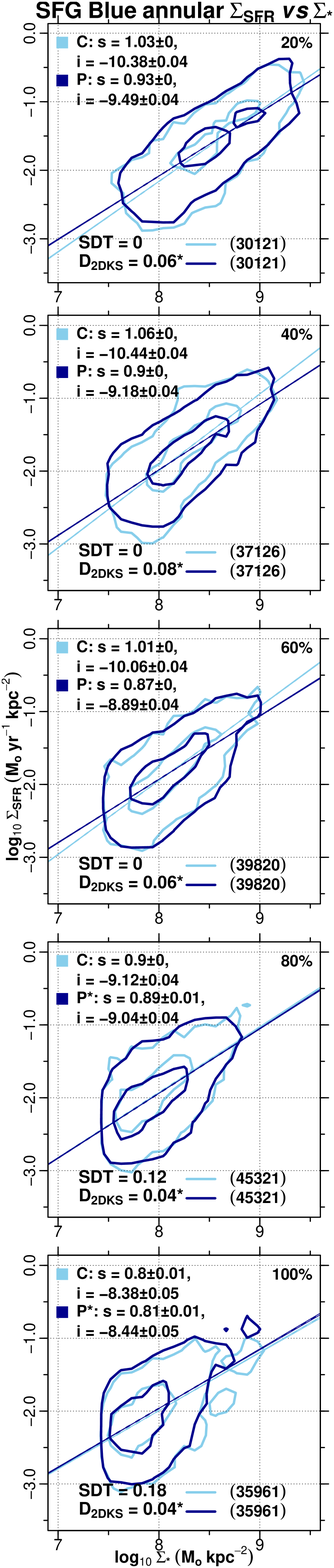}}
   \mbox{\includegraphics[width=.51665\columnwidth]{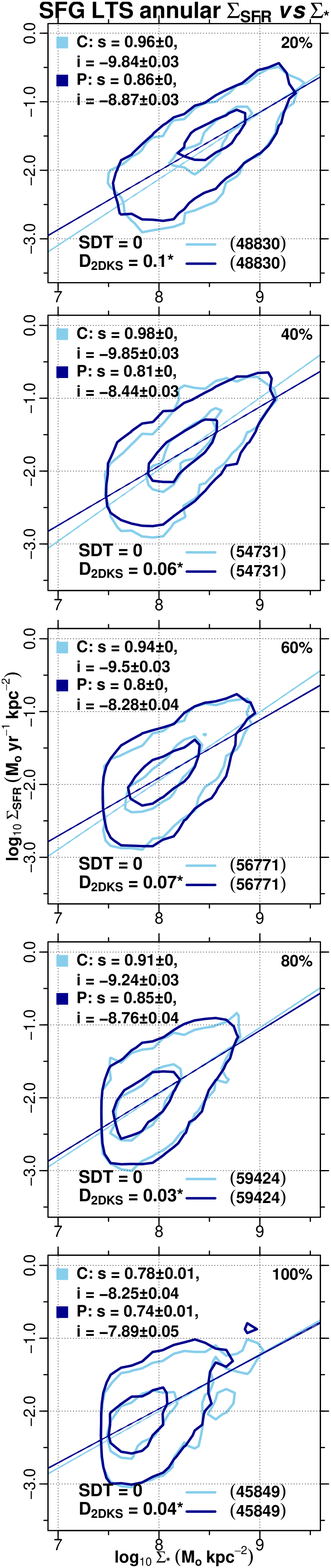}}
   \mbox{\includegraphics[width=.51665\columnwidth]{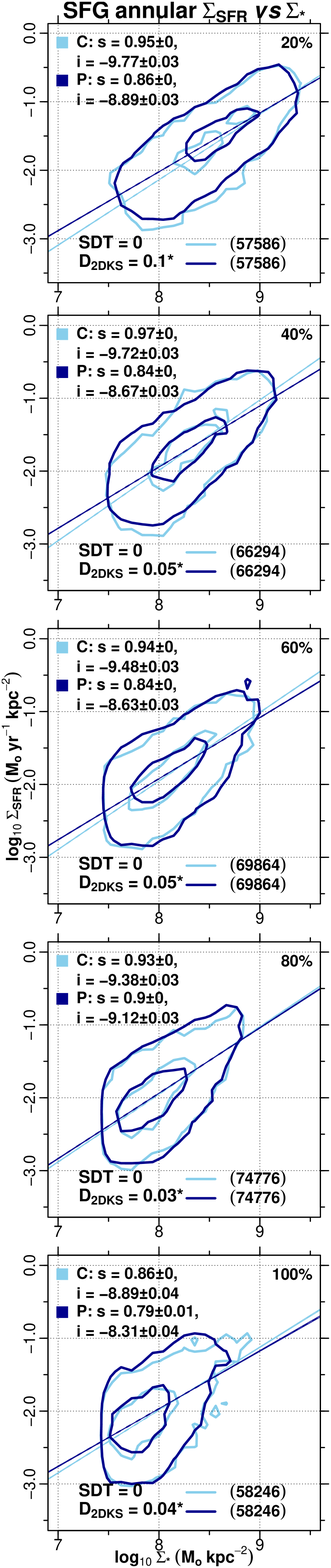}}
   \mbox{\includegraphics[width=.51665\columnwidth]{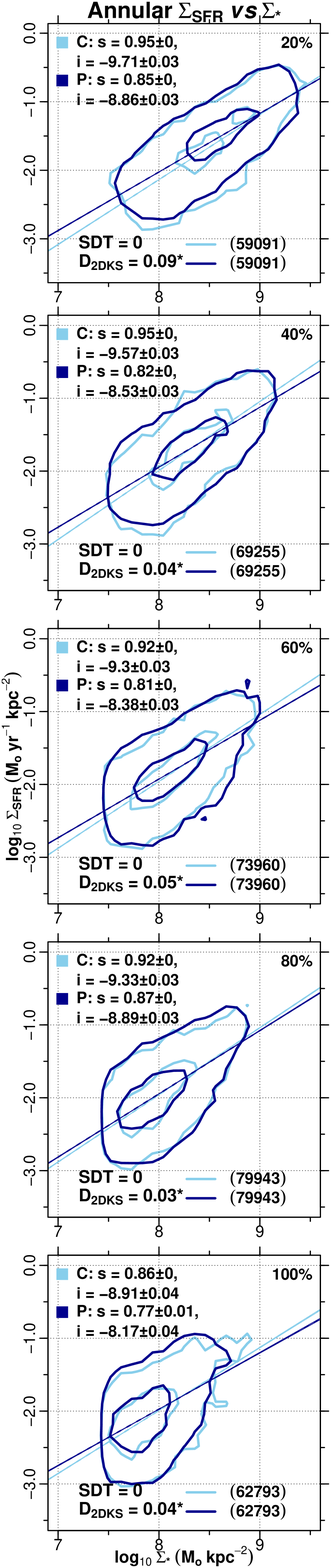}}
\caption{\scriptsize{Annular-resolved SFMSs (cont.). Same caption as above but for SFG Blue, SFG LTS, SFG and all galaxies.}}
   \label{f5} 
\end{figure*}

Galaxies with H \textsc{ii} regions, small fractions of old stars and little light concentration yield the SFMS, the tight correlation of both global M$_{*}$ and SFR. 
It has been confirmed that a local (resolved) correlation also holds for $\Sigma_{*}$ and $\mathrm{\Sigma_{SFR}}$ \citep[\textit{e.g.}][]{San13,Can16,Gon16,Hsi17,Can19,
Vul19}. Analyses which evidence the global correlation as a consequence of the resolved one have increased recently. However, \citet{EF19} and \citet{Can19} (hereafter CD-19) have shown 
that regions with low $\Sigma_{*}$ noticeably flatten the resolved relation. CD-19 exhibit that below the 7.5\,M$_{\odot}$\,kpc$^{-2}$ threshold, H$\alpha$ flux detection 
limits, apertures and other observational constraints affect the data and they warn about them in other surveys. Similarly, \citet{Hal18} 
find that galaxies and regions of others spatially below this threshold do not follow the SFMS. Due to this issue, from this Section on and in Appendices \ref{sec:app2} 
and \ref{sec:app3}, we treat only star-forming regions above the CD-19 threshold.

Table~\ref{tab:C1} lists, in annular comparison sets, the linear regression coefficients of control and perturbed samples (tA to tJ) on the SFMS plane. Specifically, 
slopes for the same predictor ($\Sigma_{*}$) across two models (control and perturbed galaxies) are compared\footnote{All SP properties overall this work are compared 
at the closest (not equal) $\Sigma_{*}$ of the star-forming regions (see Appendix~\ref{sec:app2}).}. To do so, an \emph{interaction term} between samples and $\Sigma_{*}$ 
is included in the model function. In this way, the \emph{p-value} of the interaction term represents a significance difference test (SDT) between both sample slopes. 
Importantly, sample slopes are suggested to be different when the SDT value falls below the statistical level. In each comparison set of Table~\ref{tab:C1}, slopes 
marked with * are the closest ones to the slopes resulting from combining both sample data (a topic treated afterwards, Section \ref{subsec:representatives}). Only 
SDTs with values above the statistical level are considered for this and that slope with the highest test value is marked. On the other hand, for comparisons, only 
cases of different slopes are used. Lastly, from the K-S/Peacock two-sample test, $\mathrm{D_{2DKS}}$ differences marked with * reject the null hypothesis of a parent 
distribution as the origin\footnote{Due to our large annular-subsample sizes ($n_{1}n_{2}$/($n_{1}+n_{2}$)), our $Z_{n}$ values are near the asymptotic one of the $Z$ 
statistic \citep[$Z_{\infty}$, see][]{Pea83}.}. With all this in mind, we summarize Table~\ref{tab:C1} comparisons between samples for all subsamples in Table~\ref{tab:4}.

Starting with frequencies of Different Slopes (DS), annuli having those greater than 5/10 are identified. Notice that at least in 3/5 annuli this condition is satisfied. 
That is the case of SFG Green, SFG Blue and SFGs. SFG ETS and all galaxies meet the condition in all annuli. So, in general, models of linear regression (slopes) on the 
resolved SFMS differ between control and perturbed samples. Second, from the above cases of different slopes, we find those in which the Perturbed sample has the Flattest 
slope (PF). Similarly, we distinguish those annuli in which the fraction exceeds a half. This restriction is poorly met in AGN-like and SFG Red objects (2/5, 1/5 annuli 
respectively) and is not in SFG ETSs. For the others occurs the opposite: 3/5 annuli for SFG Green, 4/5 for SFG Blue and 5/5 for the rest. This means, for both samples, 
that the $\Sigma_{*}$-$\Sigma_{\mathrm{SFR}}$ relation might be correlated with the galaxy subsample (the stellar mass concentration specifically). Finally for all annuli, 
and regardless of the subsample, the 2D distributions of control and perturbed samples point to differ in regard of their origins. All fractions of frequencies of Different 
Distribution functions (DD) are greater than a half.
 
To explore the mean trends of linear regression models of all samples on the SFMS plane, Fig.~\ref{f5} plots all star-forming regions in the annular comparison 
sets of Table~\ref{tab:C1}. In the same way, columns of Table~\ref{tab:4} allusive to Fig.~\ref{f5} list the annular frequencies result of the sample comparisons. 
DS frequencies are found in at least 3/5 annuli so, on average, linear regression slopes differ too. The previous trend of PF frequencies repeats: the lowest fractions belong 
to AGN-like, SFG Red and SFG ETS objects whereas a significant increment starts from the SFG Green subsample. As same as before, regardless of annulus and subsample, 
the 2D distributions do not share a parent one as a common origin.

Moreover, some important notes from inspecting Fig.~\ref{f5}.

\begin{itemize}
 \item{By concentric annuli: 
 \begin{enumerate}
  \item All positions of 0.1 and 0.9 contour densities clearly show a trend of both $\Sigma_{*}$ and $\Sigma_{\mathrm{SFR}}$ decreasing from the centre 
  \citep{Mar17}. 
  \item The standard deviations of both \textit{s} and \textit{i} diminish from the centre for control but remain rather constant for perturbed galaxies. Picturing 
  the stellar dynamics might distinguish perturbed galaxies as having disordered SF processes along all annuli.
  \item Excluding SFG Red objects, the highest $\mathrm{D_{2DKS}}$ differences go inwards. This might support contrasts in the transfer of gas to the centres.
 \end{enumerate}
} 
 \item{By galaxy subsamples: 
 \begin{enumerate}
  \item Compared with those of all galaxies, the density distributions of AGN-like, SFG Red, SFG ETS and SFG Green objects are quite unlike. The most dissimilar 
distributions between samples belong also to these galaxies (greater $\mathrm{D_{2DKS}}$ differences). Section~\ref{sec:dis} explores whether these facts are 
consequence of the amounts of star-forming regions that inhabit each subsample (see Table~\ref{tab:4}).
 \item As noticed already, a singular trend characterizes the flattening/steepening of the sample slopes. That trend might depend on the stellar mass 
concentration that according to the literature may distinguish our subsamples (AGN-like, SFG Red and SFG ETS objects distinguished by the 
highest $\Sigma_{*}$ values, \textit{e.g.} Appendix~\ref{sec:app2}). By finding steeper slopes for Sc-Scd/Sd-Sdm types in the disk-component SFMS, \citet{Cat17} 
propose quenched SF in Sb-Sbc more massive disks. Following this, control AGN-like, SFG Red and SFG ETS galaxies would be at a quenching stage. The same for 
perturbed SFG Green, SFG Blue, SFG LTS, and SFG galaxies. Section~\ref{sec:dis} explores whether suppressions or (re)activations of SF (quenching or rejuvenation) 
are at play.
 \end{enumerate}
} 
\end{itemize}

In sum, on the SFMS plane, star-forming regions in control and perturbed galaxies show models of linear regression that differ, inclinations of the model fits 
that depend on the galaxy subsample and 2D distributions with unlike origins.

\begin{figure*}\centering
   \mbox{\includegraphics[width=\textwidth]{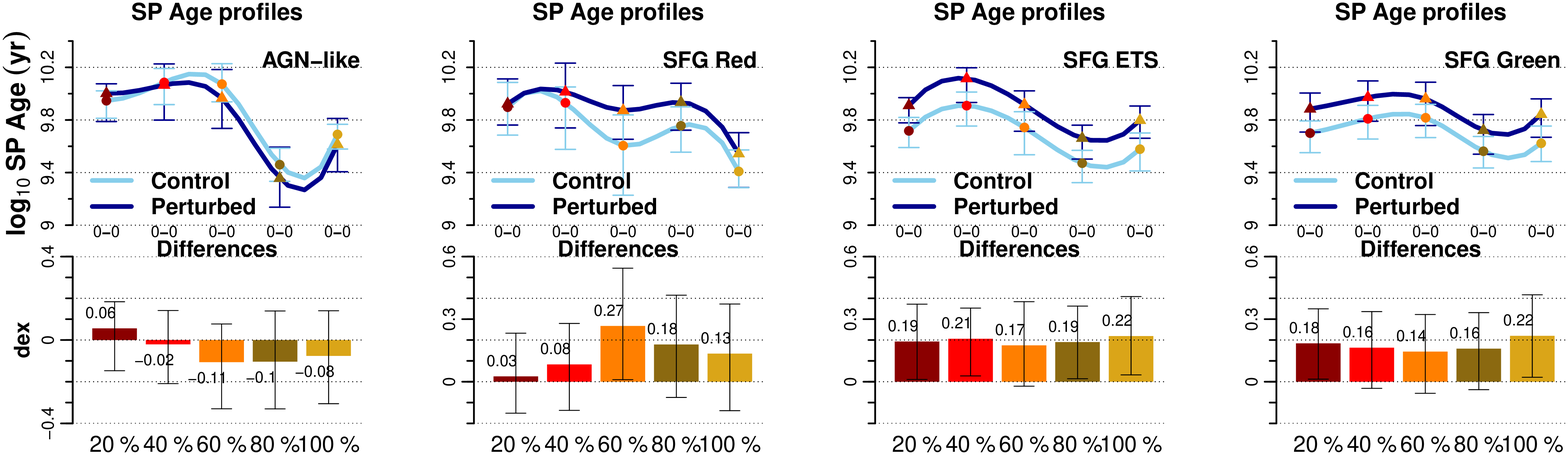}}
   \mbox{\includegraphics[width=\textwidth]{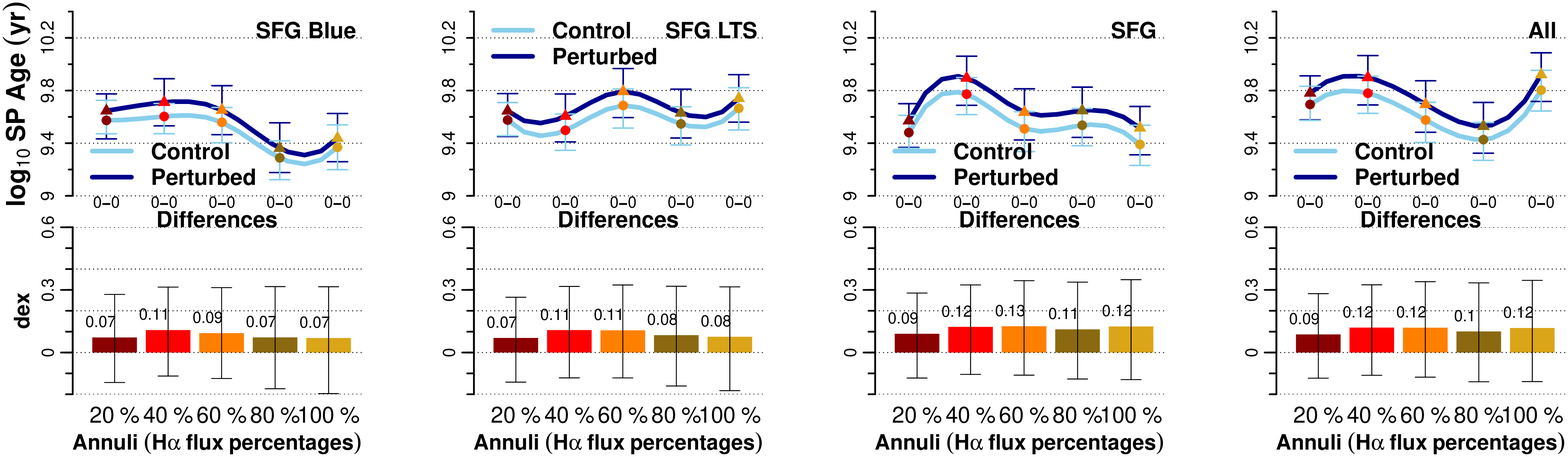}}
   \caption{\scriptsize{Annular profiles: SP age. All profiles use the amounts of annular star-forming regions per sample and subsample as in Fig.~\ref{f5}. 
Five consecutive-outward annuli denote the radial extension (see Section~\ref{subsec:annuli&profiles}). ``Differences'' (bar heights by always subtracting the control 
values from the perturbed ones) are the medians of the annular distributions of differences (differences by pairing sample spaxels which are the closest in $\mathrm{\Sigma_{*}}$). 
Bar lines depict the interquartile ranges (IQRs, 1st to 3rd) of the distributions of differences. Symbols are both sample values giving each Difference. Symbol lines 
depict the IQRs of the annular distributions. On each pair of these, AD-permutation tests are performed (likelihoods right below the profiles).}}
   \label{f6} 
\end{figure*}

\begin{figure*}
   \mbox{\includegraphics[width=\textwidth]{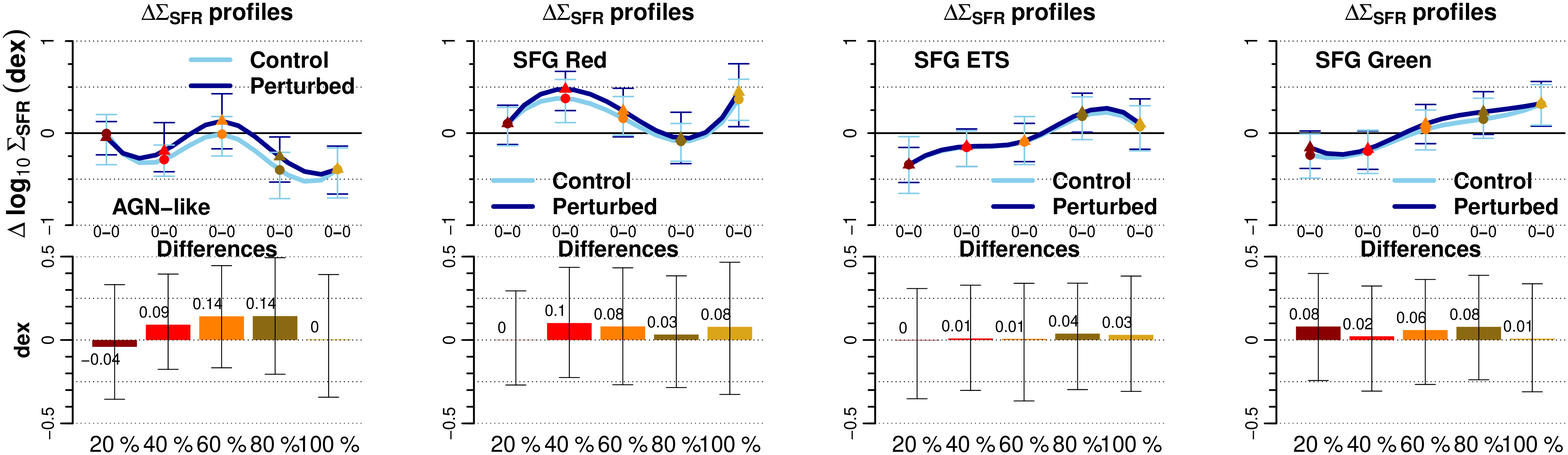}}
   \mbox{\includegraphics[width=\textwidth]{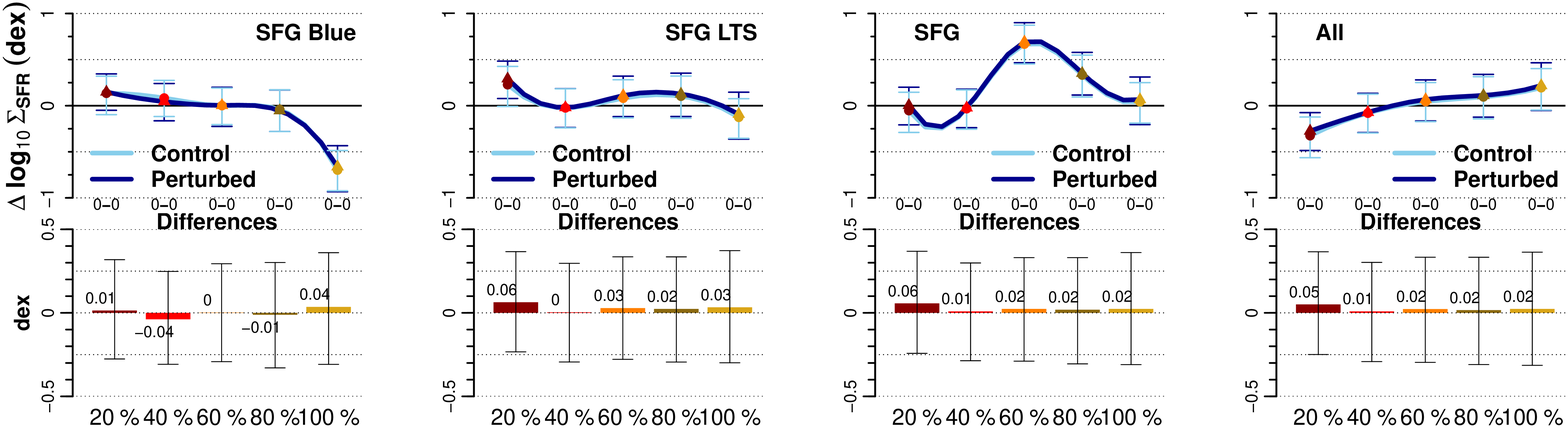}}
   \mbox{\includegraphics[width=\textwidth]{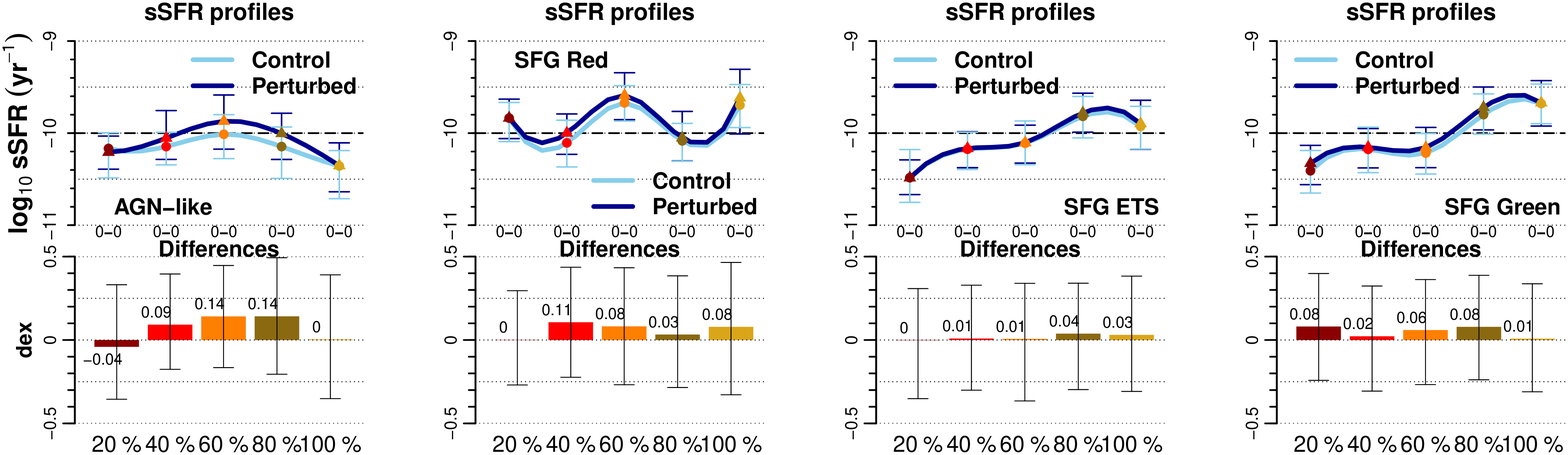}}
   \mbox{\includegraphics[width=\textwidth]{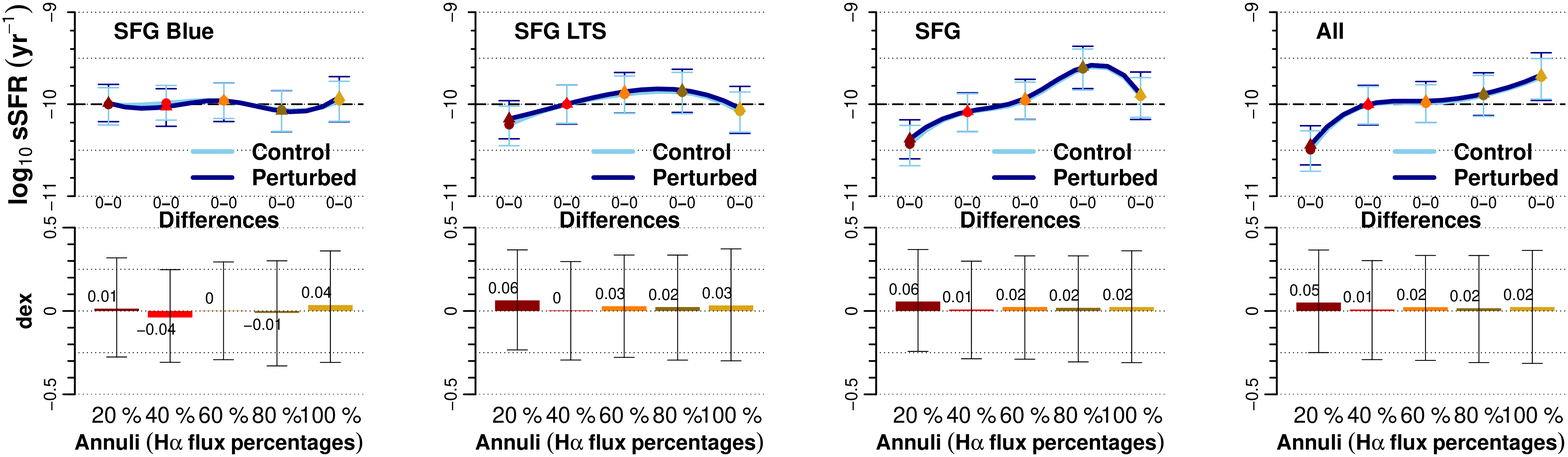}}
\caption{\scriptsize{Annular profiles: (top) $\mathrm{\Delta\,\Sigma_{SFR}}$ using the annular linear fits of all sampled regions on the SFMS 
plane, (bottom) sSFR ($\mathrm{\Sigma_{SFR}}\,\Sigma_{*}^{-1}$) where dashed lines denote the \citet{Pen10} threshold between SFGs and quiescent 
objects. Same information as in caption of Fig.~\ref{f6}.}}
   \label{f7} 
\end{figure*}

\subsection{Annular profiles}
\label{subsec:prep}

\begin{figure*}\centering
   \mbox{\includegraphics[width=\textwidth]{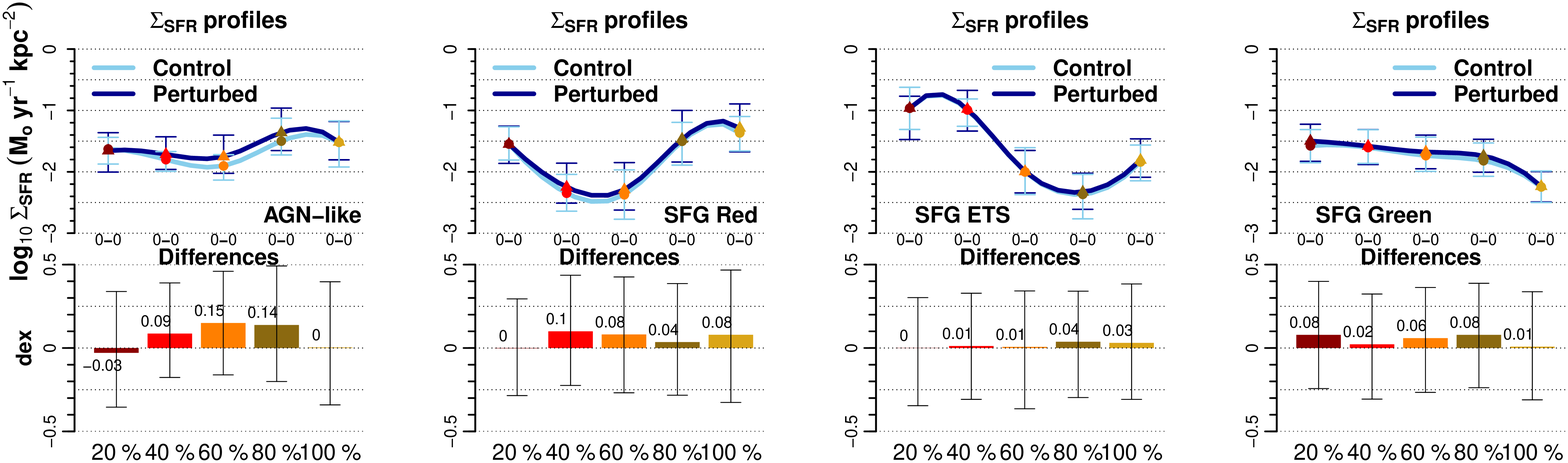}}
   \mbox{\includegraphics[width=\textwidth]{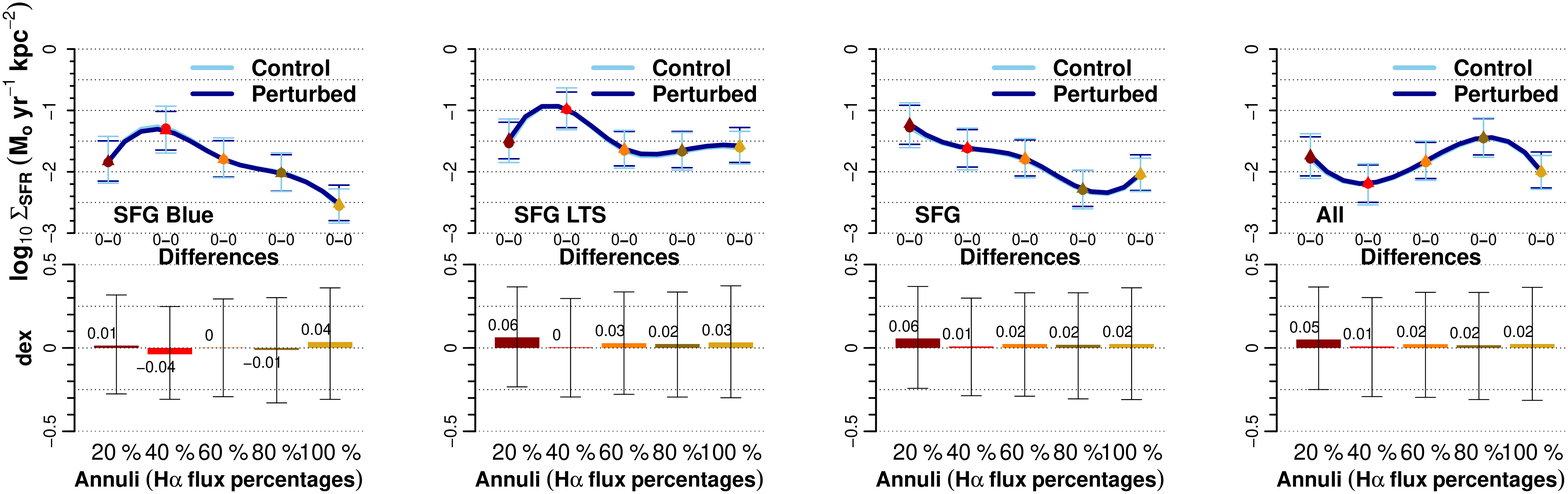}}
   \caption{\scriptsize{Annular profiles: $\mathrm{\Sigma_{SFR}}$. Same information as in caption of Fig.~\ref{f6}.}} 
   \label{f8} 
\end{figure*}

The distributions of the profiled SP properties along this work belong to all regions in the annular sets of comparisons of Table~\ref{tab:C1}. So, as in Fig. 
\ref{f5}, our profiles show the median trends across trials tA to tJ all against the control sample. Figures~\ref{f6} and \ref{f7} profile the SP age, 
$\mathrm{\Delta\,\Sigma_{SFR}}$\footnote{Properly $\Delta$\,log$_{10}$\,$\mathrm{\Sigma_{SFR}}$ (log$_{10}$\,$\mathrm{\Sigma_{SFR}}$\,$-$\,log$_{10}$\,$\mathrm{\Sigma_{SFR\,all\,sampled\,regions}}$).} and sSFR. ``Differences'' are the medians of the annular distributions of differences whilst symbols are both 
sample values giving those Differences. Line segments are the corresponding interquartile ranges (IQRs) of each respective distribution. Between samples, the distributions are compared 
by AD-permutation tests (likelihoods right below the profiles). Starting with age and excluding the AGN-like subsample, star-forming SPs are clearly older for 
perturbed samples. Differences and their IQRs are significantly extended above zero dex whereas most of the corresponding ones for AGN-like objects are well below. 
IQRs of symbols for AGN-like objects between samples clearly differ too. In perturbed galaxies of this type, SPs but the central ones tend to be younger.

Figure~\ref{f7} (top) profiles the offsets with the annular SFMSs of the SPs in all galaxies regardless of the samples. Notice clear Differences, favorable for 
perturbed samples, in AGN-like, SFG Red and SFG Green objects. Major extensions of the IQRs are above zero in 3/5 annuli and IQRs of symbols in those annuli are 
the most dissimilar. Perturbed SFG LTS, SFG and all type galaxies show favorable slight Differences only in the centres. Major extensions of IQRs of these Differences 
are clearly above zero. Total extensions of IQRs for central symbols of these perturbed subsamples are slightly biased towards higher offsets. For SFG ETS and 
SFG Blue subsamples, there are no consequential Differences ($<$10\,\%). IQRs of their Differences are more balanced around zero. Moreover, only the control 
AGN-like subsample has negative offsets (reduced) along all annuli. Reduced cases are also those of SFG ETS, SFG Green and all types: no matter the samples, 
IQRs of central symbols are totally below zero and increments towards the periphery are apparent. 

Notice the Differences and their IQRs of the sSFR (Fig.~\ref{f7} bottom) in good agreement with those of the $\mathrm{\Delta\,\Sigma_{SFR}}$. On this basis, all notes regarding the $\mathrm{\Delta\,\Sigma_{SFR}}$ 
are valid for the sSFR. Moreover, the threshold of \citet{Pen10}\footnote{It suggests that, in the local Universe, a timescale larger than log$_{10}$\,sSFR\,(yr$^{-1}$)\,$=\,-$10 
may announce the end of building the M$_{*}$ budget of a galaxy to give place to quiescence.} serves as a similar reference as the zero line for the $\mathrm{\Delta\,\Sigma_{SFR}}$. 
Based on this, two similarities and two discrepancies can be seen between the $\mathrm{\Delta\,\Sigma_{SFR}}$ and sSFR profiles. These are respectively: 1) SFG ETS, 
SFG Green and all types keep showing clear increments from the centre (SFG LTS and SFG sSFR profiles now added); 2) unexpectedly, SFG Red objects centrally increased in 
$\mathrm{\Delta\,\Sigma_{SFR}}$ and above the threshold in sSFR; 3) centres of galaxies in all subsamples but SFG Red and SFG Blue being below \citealt{Pen10} threshold 
(including their IQRs) and; 4) SFG Blue objects having a flat sSFR profile along the threshold. On the other hand, recall the younger SPs in perturbed AGN-like galaxies. 
Two facts now certainly explain this singularity: 1) regions in this subsample are increased in $\Sigma_\mathrm{{SFR}}$ and are currently more able to form stars 
\citep[\textit{e.g.}][]{San18b} and 2) perturbed AGN-like galaxies contain almost twice as many star-forming regions than their control analogues (Table~\ref{tab:2}). 

The $\mathrm{\Sigma_{SFR}}$ is next profiled in Fig.~\ref{f8}. The agreement of the Differences and their IQRs between $\mathrm{\Delta\,\Sigma_{SFR}}$ and sSFR 
apparently persists in the $\mathrm{\Sigma_{SFR}}$. This is a mere coincidence and has nothing to do with a common property ($\mathrm{\Sigma_{SFR}}$) influencing 
the computations.\footnote{Footnote 9 and the three properties profiled differently (unlike shapes) argue against the fact of Differences belonging to only one 
particular region.} The previous allows us to state, as in the case of $\mathrm{\Delta\,\Sigma_{SFR}}$ and sSFR, that the favorable Differences for perturbed samples 
are: 1) mostly dominant in AGN-like, SFG Red and SFG Green subsamples (3/5 annuli); 2) little, but evidently increased only centrally in SFG LTS, SFG and all 
subsamples and; 3) very likely, neither increased nor reduced in SFG ETS and SFG Blue ones. Notice, in addition, that the statistical tests on the annular distributions 
result in the lowest likelihoods regardless of the property and subsample.

In general, the SFR properties of regions in perturbed galaxies are practically never inferior. This is despite the fact that these regions are older.

\section{Discussion}
\label{sec:dis}

\subsection{Amounts of regions and their SFMS distributions}
\label{subsec:amounts}

\begin{figure*}\centering
   \mbox{\includegraphics[width=.51665\columnwidth]{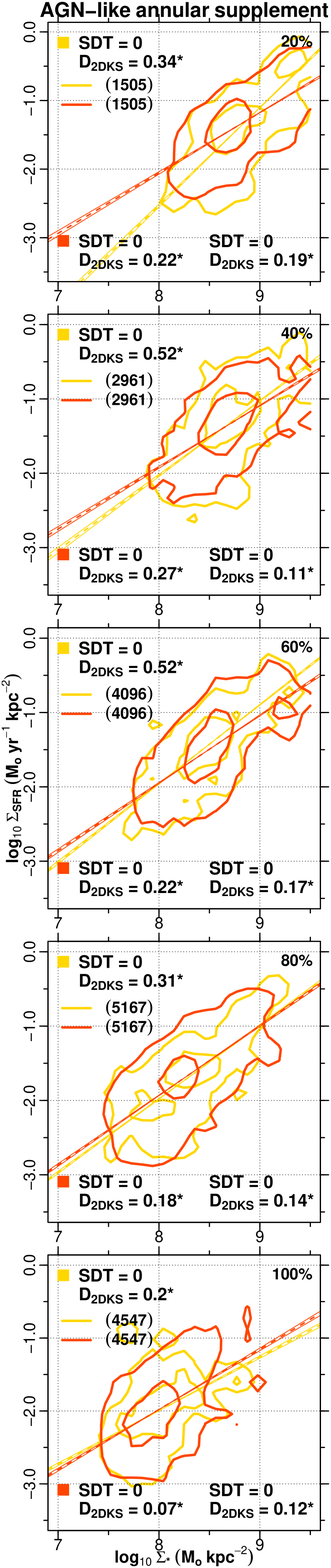}}
   \mbox{\includegraphics[width=.51665\columnwidth]{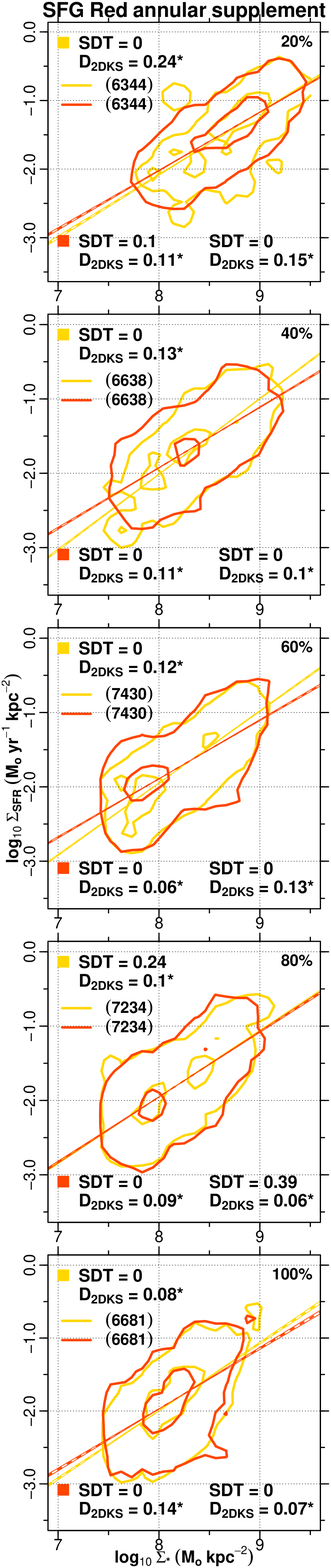}}
   \mbox{\includegraphics[width=.51665\columnwidth]{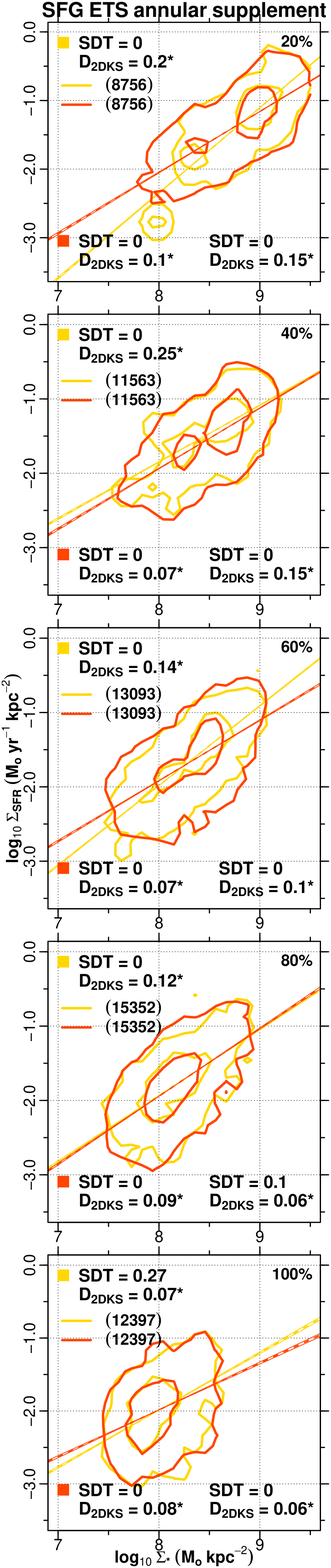}}
   \mbox{\includegraphics[width=.51665\columnwidth]{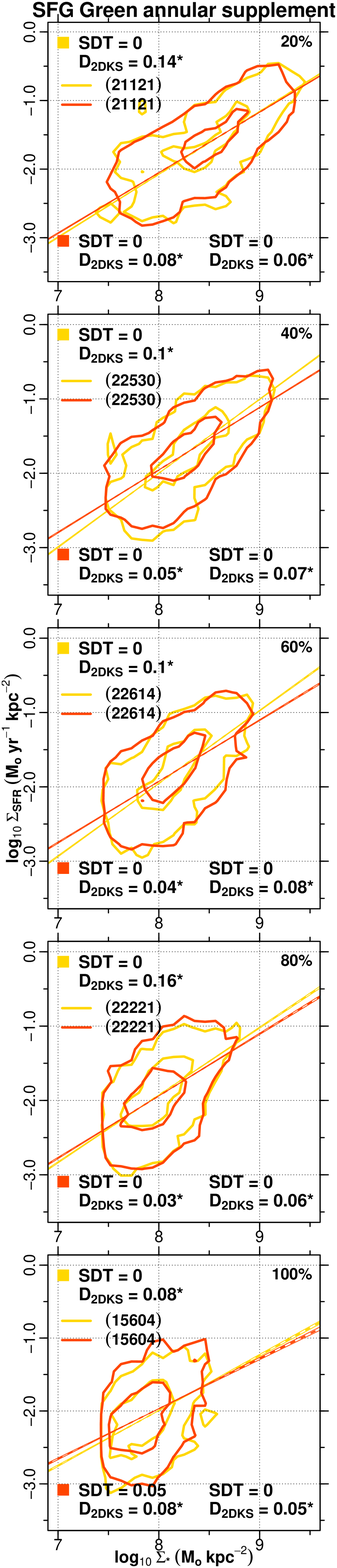}}
\caption{\scriptsize{Supplemental, annular-resolved SFMSs. Perturbed star-forming regions (red) are paired with their control ones (gold) at the closest $\mathrm{\Sigma}_{*}$. 
All supplemental star-forming regions in each comparison set (\textit{i.e.} trials tA to tJ each one with its respective control sample regions, see Appendix~\ref{sec:app3}) 
are considered. From left to right: Kernel density estimations (0.1 and 0.9 contour densities from outside-in) of the supplemental star-forming regions corresponding to 
AGN-like, SFG Red, SFG ETS and SFG Green galaxies. Models for all annular sequences are statistically significant (error intervals are drawn for all fits). SDTs and 
$\mathrm{D_{2DKS}}$ differences (asterisks indicate both distribution functions coming from a different parent one) following colour squares (gold, control; red, 
perturbed) give the test results between regions (Fig.~\ref{f5} 1st part) and their supplemental ones within the same sample. SDTs and $\mathrm{D_{2DKS}}$ differences 
at the bottom-right give the results between supplemental regions only.}}
   \label{f9} 
\end{figure*}

In Section~\ref{subsec:mssf_1} we show that the density distributions on the SFMS plane of four subsamples: AGN-like, SFG Red, SFG ETS and SFG Green, are dissimilar 
compared with those of all subsamples. Between samples, $\mathrm{D_{2DKS}}$ differences are also greater among these subsamples than those among the other types. 
Table~\ref{tab:4} lists the totals of star-forming regions used in each subsample. Notice a significant unbalance for AGN-like, SFG Red, SFG ETS and SFG Green 
subsamples with respect to the rest. To get a better understanding of a possible dependence of the SFMS distributions (the density distributions on the SFMS plane) on the amount 
of star-forming regions, in Fig.~\ref{f9} we show the supplemental SFMSs. These help to characterize regions from all galaxies except those corresponding to each one of the 
four subsamples, \textit{i.e.}, Fig.~\ref{f5} 1st part. Notice that the selection of these 
supplemental regions is done at the closest $\Sigma_{*}$ values. These supplemental regions are distinguished either within control (gold contours) or within 
all the perturbed samples (red contours). The reasoning is the following: if the parameters derived from these supplemental distributions are alike to those 
of Fig.~\ref{f5} 1st part, we would prove a dependence on the amount of regions. The contrary would mean that the SF processes (which characterize the regions and 
the supplemental ones) are different.

It turns out from Fig.~\ref{f9}, that SDTs point to equal slopes in only four annuli: two for SFG Red (20 and 80\,\% annuli, red and gold squares) and one for 
SFG ETS and SFG Green each (100 and 100\,\% annuli, gold and red squares, respectively). Regarding $\mathrm{D_{2DKS}}$ differences, all suggest that the density 
functions being compared do not have their respective origins in a common parent distribution. In sum, there are no similitudes between regions and their supplemental 
ones, at the closest $\Sigma_{*}$ values and at the same number of regions for the perturbed samples and the control one. The dissimilarities between density 
distributions mentioned above do not depend neither on the amounts nor on the stellar mass concentration of the regions. On the other hand, by looking at the parameters 
comparing both supplemental regions between samples (bottom-right in all panels of Fig.~\ref{f9}), in 18/20 annuli, the SDTs suggest different 
slopes. From these frequencies, the fit corresponding to perturbed samples (red) is the flattest in 16 cases ($\mathrm{D_{2DKS}}$ differences also point to uncommon 
origins). Therefore the linear fits, either flatter or steeper, do not depend on the $\Sigma_{*}$ of the regions but on some other property characterizing the 
galaxy types defining our subsamples. A possible explanation involving the M$_{*}$ along with the perturbation parameter is treated in Section~\ref{subsec:dependence}.

The one certainty is that most of the slopes for perturbed objects are flatter, than the corresponding ones for control objects, at the $\Sigma_{*}$ values closest 
to those ones found in AGN-like, SFG Red, SFG ETS and SFG Green types. Concluding, the dominant trend for the galaxies in this study (\textit{i.e.} the majority 
of regions, see Table~\ref{tab:4}) is that perturbed ones show lowered intensities of the SFR with increasing stellar mass density.

\subsection{Suppressions and/or (re)activations of SF}
\label{subsec:quench-rejuv}

Section~\ref{subsec:mssf_1} shows slopes with a peculiar behaviour on the SFMS plane. These are flatter and steeper, for control and perturbed galaxies respectively, 
in AGN-like, SFG Red and SFG ETS objects. Then the slopes switch, to either flatter or steeper, in SFG Green, SFG Blue, SFG LTS, SFG and all objects. Since the former 
group of subsamples is distinguished by higher concentration of stellar mass (see Appendix~\ref{sec:app2}), this peculiar slope behaviour might depend on $\Sigma_{*}$. 
In the scenario proposed by \citet{Cat17}, control AGN-like, SFG Red and SFG ETS galaxies would be at a quenching stage (same for those in perturbed SFG Green, SFG Blue, SFG LTS and 
SFG subsamples).

However, the results in Section~\ref{subsec:amounts} suggest that this may not be the case. Steeper slopes for regions in perturbed galaxies do not repeat at stellar mass densities which are 
the closest to those found in AGN-like, SFG Red, SFG ETS and even SFG Green objects. To clarify on this matter, the SFR ($\mathrm{\Delta\,\Sigma_{SFR}}$, sSFR and $\mathrm{\Sigma_{SFR}}$) 
profiles (Figs.~\ref{f7} and \ref{f8}) are reviewed for traces of suppression (quenching) and/or (re)activation (rejuvenation) of SF. Along each property scale, we 
contrast the positions of central values relative to those ones of the rest annuli. This approach is justified since changes in central regions 
(either enhancement or suppression) dominate the regulation of SF \citep{Ell18a}.

Starting with the $\mathrm{\Delta\,\Sigma_{SFR}}$, SFG Blue and SFG LTS galaxies show central offsets above the rest annuli. In AGN-like objects, the central offset 
exceeds three of the rest annuli. SFG Red galaxies have its central offset below three of the rest annuli. The rest galaxy types show evident increments from the 
centre. Regarding sSFRs, the SFG Red subsample central annulus appears diminished by two annuli. Though quite aligned with all the rest, the central annulus in SFG 
Blue objects is diminished by two annuli. The central sSFRs and their IQRs in the rest subsamples evidence increments as going outwards (not at all in 
AGN-like objects due to the periphery). Lastly, the central $\mathrm{\Sigma_{SFR}}$ is diminished against the periphery in AGN-like and SFG Red subsamples. The 
same may be for all subsamples together. In SFG Blue and SFG LTS types, the central intensity is just diminished against that one of the next annulus. It is in 
SFG ETS, SFG Green and SFG objects where the centre is evidently dominant. Summarizing, the three profiles show diminished central values (against one annuli 
at least) in AGN-like, SFG Red and all subsamples. These three types of galaxies are then the clearest ones suffering from 
quenching.

Though AGN-like types are clear candidates for quenching, some attributes distinguish the perturbed objects. Their age profile indicates younger SPs along all 
annuli but the central one. In the middle (40, 60 and 80\,\% annuli), perturbed samples clearly advantage the control one in the three SFR profiles just discussed. 
In average, numbers of star-forming regions for perturbed samples double control frequencies. This all suggests rejuvenation of the mid-annular SPs in perturbed 
AGN-like objects.

Finally, the favorable central Differences (Figs.~\ref{f7} and \ref{f8}) for perturbed SFG LTS, SFG and, specially, all galaxies, place control objects 
as the closest candidates for quenching. An interesting fact is why perturbed galaxies mostly exhibit flatter slopes on the SFMS plane 
(Table~\ref{tab:4} and Section~\ref{subsec:amounts}). Perhaps the influence of stellar mass density is somehow weakened. Something else might be regulating the 
SF in perturbed galaxies.

\subsection{A correlation between SF and tidal effects}
\label{subsec:dependence}

\citet{Pen10} estimate the mean local density of galaxies at different redshifts. There are no detectable differences in SFMS and mean sSFR for SFGs between the 
lowest and highest density Qs. \citet{Ell18a} find a similar result. By doing a dynamical analysis, \citet{Er16} compile secure X-ray-detected galaxies to explore 
the SFMS too. At $0.15<\,z\,<0.50$ they find a flattening caused by red-disk dominated galaxies. Notice that the galactic vicinities in these analyses are expressed 
by density fields whereas our approach sets the need of ensuring physically related objects, \textit{i.e.} tidally-related, particularly. As a result, on the SFMS 
plane, most of the star-forming regions in perturbed galaxies reproduce flatter linear models than those ones for regions in control objects (Section 
\ref{subsec:mssf_1}). This $\mathrm{\Sigma_{SFR}}$ reduction with $\mathrm{\Sigma_{*}}$ points to quenched SF since, indeed, SPs in the perturbed objects are mostly 
older. However, this assumption is questionable since regions in perturbed objects are practically never inferior in the SFR profiled 
properties (Section~\ref{subsec:prep}). Contrasts are clearly favorable for perturbed AGN-like, SFG Red and SFG Green galaxies. For perturbed SFG LTS, SFG and all 
galaxies, contrasts are also favorable, though only in the centres. Though the SFR profiled properties confirm that a SF suppression has likely started in the centres (Section \ref{subsec:quench-rejuv}), 
this is regardless of either control or perturbed samples. Moreover, SPs in perturbed AGN-like objects are younger than SPs in their control analogues 
and the SFR profiled properties of the former are evidently higher. In sum, the flattening on the SFMS plane may not be due to SF quenching but due to a less 
dependent $\mathrm{\Sigma_{SFR}}$ on $\Sigma_{*}$. Tidal effects might be contributing to drive SF in these regions. This hypothesis would explain the (sometimes 
little but perceptible) contrasts in the SFR properties.
 
\begin{figure}\centering
   \mbox{\includegraphics[width=1.015625\columnwidth]{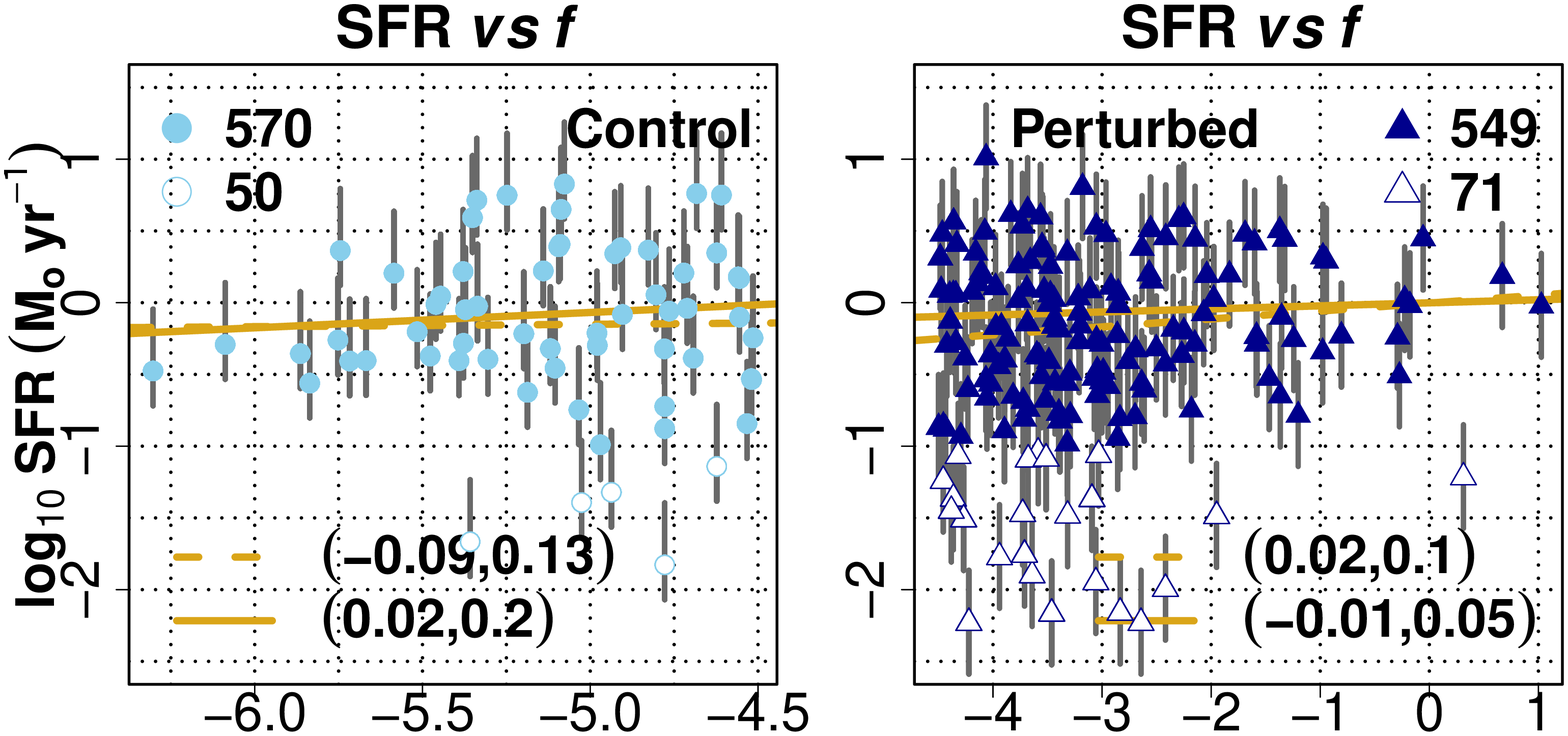}}
   \mbox{\includegraphics[width=1.015625\columnwidth]{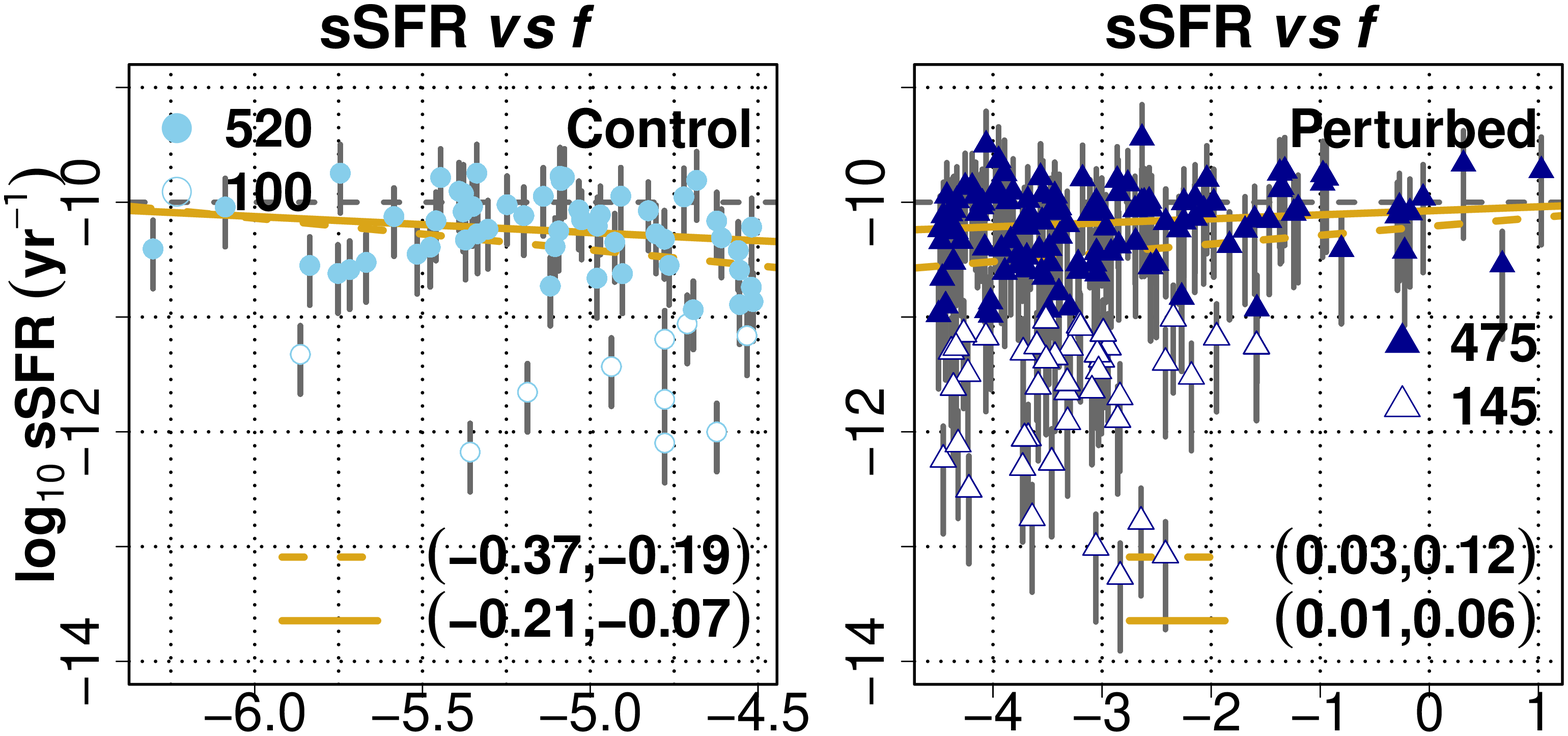}}
   \mbox{\includegraphics[width=1.015625\columnwidth]{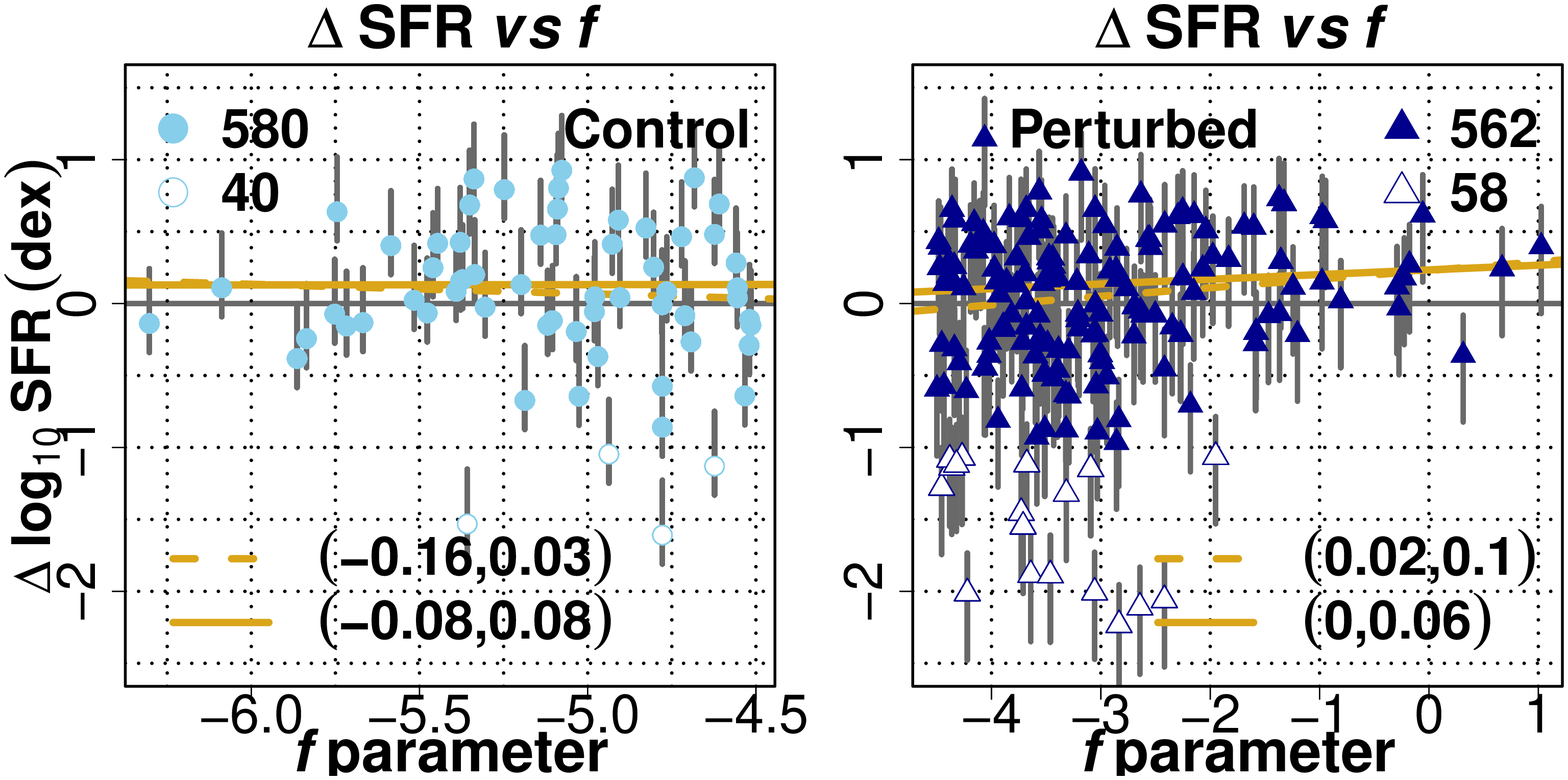}}
\caption{\scriptsize{The SFR properties against the \textit{f} parameter for control (Control) and perturbed (Perturbed, tA to tJ) samples. Black-dashed lines for the sSFR as the \citet{Pen10} 
threshold between SFGs and quiescent objects. The SFR offset ($\Delta$\,SFR) is computed from a linear fit (SFMS) on the data of all samples. Line segments of 
points as the IQRs of each property distribution in each sample. Numbers after filled and empty symbols are the frequencies of points above and below, respectively, 
the thresholds of $-$1, for SFR and $\Delta$\,SFR, and $-$11, for sSFR. Gold lines are robust linear model fits, for all points (dashed), and for filled ones 
(solid). Intervals at 95\,\% confidence level (bottom) enclose the true values of the model slopes (gold fits). Complementary statistical results in Table 
\ref{tab:5}.}}
\label{f10} 
\end{figure}

A possible correlation with tidal effects is then explored. We first combined our resolved results alongside the \textit{f} parameter global measurements (defined 
in Section~\ref{subsec:f}). When pairing the spaxels at a specific value of $\Sigma_{*}$ (the annular median of a galaxy in any sample or the median difference between 
both annular sets of a galaxy in a sample and the rest galaxies in the other sample) there are always cases of galaxies in a sample with more than one $\Sigma_{*}$ 
value as the closest to another one of another galaxy in the other sample. This issue makes \textit{f} parameters to have more than one resolved property value at 
the same time, having, as a consequence, perceptible effects on the correlation. Notice, in contrast, that our resolved-to-resolved correlations are free from this 
issue. Figure~\ref{f10} uses therefore integrated SFR properties of the samples at tight M$_{*}$ distributions (see Fig.~\ref{f1}). As we previously did (Figs.~\ref{f5} 
and \ref{f9}), the average trend is shown for perturbed samples by plotting them altogether alongside versions of the control set. From robust model regression, 95\,\% 
confidence intervals of the resultant slopes (gold lines) are included. Moreover, Table~\ref{tab:5} lists the summary of the 
linear model and correlation results. By looking at the gold dashed lines in Fig.~\ref{f10}, notice positive intervals for perturbed samples and negative lower ends 
for the control one. From the corresponding slopes, ``dashed'' rows in Table~\ref{tab:5}, perturbed samples have both linear model and correlation p-values below the 
statistical level. There seems to be a little but detectable increment of the SFR properties in perturbed galaxies with \textit{f}. By definition, galaxies in the control 
sample are not experiencing gravitational torques that disturb their secular growth of stellar mass. Intervals and linear fits should be roughly balanced and flat,
meaning that the tidal perturbation scale for the control sample is rather unimportant. If we now look on how the perturbed galaxies are distributed, we notice scarcities 
of high values for high parameters along with abundances of low values for low parameters.\footnote{These profusions mostly consist of AGN-like and early type objects 
the ones characterized by the oldest SP median ages.} To measure the effect of these unbalances, we recalculate the slope confidence interval and correlation coefficient 
for points\,$\geq$\,log$_{10}\,-$1 M$_{\odot}$\,yr$^{-1}$, $\geq$\,log$_{10}\,-$11 yr$^{-1}$ and $\geq\,-$1 dex, for the SFR, sSFR and $\Delta$\,SFR respectively (gold 
solid lines in Fig.~\ref{f10} and ``solid'' rows in Table \ref{tab:5}). We find, for the perturbed samples, that though the intervals are reduced, slopes keep on being 
most likely positive and that the slope and correlation coefficient fail now on being significant only for the SFR. For the control sample, the only property that strongly 
varies is precisely the SFR. Since sSFR and $\Delta$\,SFR increments continue for perturbed galaxies, we may conclude that tidal perturbations play somehow a role as a 
SF driver.

In contrast, \citet{Cat17} find M$_{*}$ as the absolute driver. In this regard, we recall the trend of steeper slopes (on the SFMS plane) for regions in perturbed 
AGN-like, SFG Red and SFG ETS subsamples. The reason may certainly involve the characterisctic M$_{*}$ values for galaxies of these types. Perhaps M$_{*}$ is too 
dominant so any effect of tidal perturbations (flattening included) is wiped out. We explore the M$_{*}$-\textit{f} parameter distributions for only perturbed 
galaxies. The ones in these three types are biased towards higher stellar masses and towards lower perturbation parameters. The opposite occurs for galaxies in 
the rest subsamples. Unfortunately, the anticorrelations result of comparing both global measurements are not statistically significant.

\subsection{An effect of perturbed galaxies on the SFMS}
\label{subsec:representatives}

\begin{table}
   \setlength{\tabcolsep}{0.1\tabcolsep}
 \begin{minipage}{\columnwidth}
\caption{\scriptsize{Summary of results for the SFR properties against the \textit{f} parameter. Robust linear model slope \textit{s}, its standard error \textit{e} 
and model \textit{p-value}. Pearson correlation coefficient ($\mathrm{r_{P}}$) and related \textit{p-value}. ``dashed'' and ``solid'' rows in reference to the linear 
fits of Fig.~\ref{f10} (dashed and solid gold lines).}
\label{tab:5}}
  \centering
 \begin{scriptsize}
 \begin{tabular}{l@{\hspace{0.7\tabcolsep}}c@{\hspace{0.7\tabcolsep}}c@{\hspace{0.7\tabcolsep}}c@{\hspace{0.7\tabcolsep}}c@{\hspace{0.7\tabcolsep}}c@{\hspace{0.7\tabcolsep}}c@{\hspace{0.7\tabcolsep}}c@{\hspace{0.7\tabcolsep}}c@{\hspace{0.7\tabcolsep}}c@{\hspace{0.7\tabcolsep}}c@{\hspace{0.7\tabcolsep}}c}
 \hline
       &                             \multicolumn{5}{c}{Control}                                 &&                             \multicolumn{5}{c}{Perturbed}                               \\
       &\textit{s}       &\textit{e}       &\textit{p-value} &$\mathrm{r_{P}}$ &\textit{p-value} &&\textit{s}       &\textit{e}       &\textit{p-value} &$\mathrm{r_{P}}$ &\textit{p-value} \\
\cline{2-6}\cline{8-12}                                                                                                                                                                     \\                                                                                                                
       &                                                           \multicolumn{11}{c}{log$_{10}$\,SFR (M$_{\odot}$\,yr$^{-1}$)}                                                            \\
dashed &\phantom{0}0.017 &\phantom{0}0.056 &\phantom{0}0.760 &$-$0.007         &\phantom{0}0.867 &&\phantom{0}0.055 &\phantom{0}0.021 &\phantom{0}0.006 &\phantom{0}0.114 &\phantom{0}0.004 \\
solid  &\phantom{0}0.107 &\phantom{0}0.047 &\phantom{0}0.021 &\phantom{0}0.082 &\phantom{0}0.050 &&\phantom{0}0.021 &\phantom{0}0.017 &\phantom{0}0.210 &\phantom{0}0.061 &\phantom{0}0.155 \\[1ex]
       &                                                                 \multicolumn{11}{c}{log$_{10}$\,sSFR (yr$^{-1}$)}                                                                  \\
dashed &$-$0.276         &\phantom{0}0.046 &\phantom{0}0     &$-$0.213         &\phantom{0}0     &&\phantom{0}0.077 &\phantom{0}0.022 &\phantom{0}0.001 &\phantom{0}0.136 &\phantom{0}0.001 \\
solid  &$-$0.140         &\phantom{0}0.034 &\phantom{0}0     &$-$0.201         &\phantom{0}0     &&\phantom{0}0.035 &\phantom{0}0.012 &\phantom{0}0.005 &\phantom{0}0.134 &\phantom{0}0.004 \\[1ex]
       &                                                                \multicolumn{11}{c}{$\Delta$ log$_{10}$\,SFR (dex)}                                                                 \\
dashed &$-$0.066         &\phantom{0}0.050 &\phantom{0}0.180 &$-$0.086         &\phantom{0}0.033 &&\phantom{0}0.060 &\phantom{0}0.019 &\phantom{0}0.002 &\phantom{0}0.129 &\phantom{0}0.001 \\
solid  &\phantom{0}0.002 &\phantom{0}0.042 &\phantom{0}0.964 &$-$0.025         &\phantom{0}0.550 &&\phantom{0}0.033 &\phantom{0}0.016 &\phantom{0}0.035 &\phantom{0}0.090 &\phantom{0}0.033 \\[1ex]
\hline\\
 \end{tabular}
 \end{scriptsize}
 \end{minipage}
 \end{table}

\begin{table*}
 \begin{minipage}{\textwidth}
\caption{\scriptsize{Samples hosting the best representative linear regression models (slopes) of those ones when compiling the data of both samples in each respective 
annular comparison set (see Table~\ref{tab:C1}). Perturbed samples are labeled as usually (tA to tJ). The control one (C) has an additional letter which indicates to 
what perturbed sample is compared to. Small fonts indicate that the slopes are the flattest ones of each set. ``f1'' gives the frequencies of best representatives. ``f2'' 
gives the frequencies where the sample slope is the flattest one (small fonts) and ``f3'' tells, from f2, the cases that belong to the perturbed sample.}
 \label{tab:6}}
 \centering
 \begin{scriptsize}
 \begin{tabular}{@{\hspace{0.065\tabcolsep}}l@{\hspace{0.065\tabcolsep}}c@{\hspace{0.75\tabcolsep}}c@{\hspace{0.75\tabcolsep}}c@{\hspace{0.75\tabcolsep}}c@{\hspace{0.75\tabcolsep}}c@{\hspace{0.75\tabcolsep}}c@{\hspace{0.25\tabcolsep}}c@{\hspace{0.25\tabcolsep}}c}
\hline
                                                                                                                        \multicolumn{6}{c}{Annuli (H$\alpha$ flux percentages)}                                                                                                                                                                                           &     &     &     \\
                &20\%                                           &40\%                                        &60\%                                                                      &80\%                                                                                                 &100\%                                                                  &f1   &f2   &f3   \\
\hline
\tiny{AGN-like} &tD                                             &tB tD tI tJ                                 &tA \texttt{tB} tC \texttt{tD} tE tF tG tH \texttt{tI} \texttt{tJ}         &tA \texttt{tB} \texttt{CD} tE \texttt{tG} CI CJ                                                      &\texttt{tA} \texttt{tB} \texttt{tD} \texttt{tE} \texttt{tH} \texttt{tI}&28/50&13/28&12/13\\[1ex]
\tiny{SFG Red}  &tA tB \texttt{CC} tD tE \texttt{tG} \texttt{tH}&CA CB CC CD CE                              &tB tC tD tE tG \texttt{tH} \texttt{tI} tJ                                 &tG \texttt{tH} \texttt{tI}                                                                           &\texttt{tG} tJ                                                         &25/50& 8/25&  7/8\\[1ex]
\tiny{SFG ETS}  &\texttt{CB} \texttt{CE} CG \texttt{CI}         &$\ldots$                                    &\texttt{CG}                                                               &\texttt{CA} \texttt{CB} CC \texttt{CD} \texttt{CE} \texttt{CF} tG \texttt{CH} \texttt{CI} \texttt{CJ}&$\ldots$                                                               &15/50&12/15& 0/12\\[1ex]
\tiny{SFG Green}&\texttt{tD} \texttt{CG}                        &\texttt{tB} CC \texttt{tD} \texttt{CE} CH CJ&tA \texttt{tB} \texttt{tE} \texttt{tF} \texttt{tH} \texttt{CI} \texttt{tJ}&tA \texttt{tB} tE tF tH \texttt{tI} tJ                                                               &\texttt{tB} \texttt{tE} \texttt{tF} \texttt{tH} \texttt{tJ}            &27/50&18/27&15/18\\[1ex]
\tiny{SFG Blue} &CA CC \texttt{tG}                              &$\ldots$                                    &CC                                                                        &\texttt{tA} \texttt{tB} \texttt{CC} CD tE \texttt{tF} \texttt{tG} \texttt{tH} tI \texttt{tJ}         &tA tB \texttt{tD} \texttt{tE} \texttt{tG} tH tI tJ                     &22/50&11/22&10/11\\[1ex]
\tiny{SFG LTS}  &CA CE                                          &$\ldots$                                    &$\ldots$                                                                  &\texttt{tA} \texttt{tC} tE \texttt{tF}                                                               &tA \texttt{tB} \texttt{tD} tE tJ                                       &11/50& 5/11&  5/5\\[1ex]
\tiny{SFG}      &CA CE                                          &CA                                          &\texttt{tA} \texttt{tE}                                                   &\texttt{tA} \texttt{tB} \texttt{tC} tE \texttt{tF} \texttt{tH} \texttt{tI} \texttt{tJ}               &\texttt{tA} \texttt{tB} \texttt{tD} tE \texttt{tJ}                     &18/50&13/18&13/13\\[1ex]
\tiny{all}      &CA CE                                          &CA                                          &\texttt{tE}                                                               &\texttt{tA} \texttt{tC} \texttt{tE} \texttt{tF} \texttt{tI}                                          &\texttt{tA} \texttt{tB} tC \texttt{tD} \texttt{tE} \texttt{tJ}         &15/50&11/15&11/11\\[1ex]
f1              &23/80                                          &17/80                                       &30/80                                                                     &54/80                                                                                                &37/80                                                                  &     &     &     \\
f2              &9/23                                           &3/17                                        &16/30                                                                     &37/54                                                                                                &26/37                                                                  &     &     &     \\
f3              &4/9                                            &2/3                                         &14/16                                                                     &27/37                                                                                                &26/26                                                                  &     &     &     \\
\hline\\
 \end{tabular}
 \end{scriptsize}
 \end{minipage}
 \end{table*}

\citet{Can16} report a SFMS of $s\,=\,$0.72, $i\,=\,-$7.95 with $\mathrm{\sigma}\,=$\,0.16\,dex. As same as CD-19, who obtain $s\,=\,$0.94, $i\,=\,-$10.00 and 
$\mathrm{\sigma}\,=$\,0.27\,dex, we use the OLS method and their threshold in $\Sigma_{*}$. We neither intend to compare the global and local SFMSs within our 
data, nor to explore the flattening below the threshold. Therefore, we do not bin the data as CD-19 do. All star-forming regions in all samples give 
$s\,=\,$0.84, $i\,=\,-$8.69 with $\mathrm{\sigma}\,=$\,0.34\,dex. Regions in the control sample give $s\,=\,$0.87, $i\,=\,-$8.94 with $\mathrm{\sigma}\,=$\,0.33\,dex 
whereas regions in all perturbed samples give $s\,=\,$0.81, $i\,=\,-$8.44 with $\mathrm{\sigma}\,=$\,0.35\,dex. The following remarks emerge from this. First, 
the sequences reported above still come from pairing the star-forming regions at the closest $\Sigma_{*}$ (see Fig.~\ref{f5} for all galaxies). Second, our 
scatters are as high as the ones reported by \citet{Mar17}, \citet{Hal18} and \citet{Vul19}. And last, we look for any effect perturbed galaxies may cause on 
the SFMS. 

The high scatters are due to systematic diversities and subgalactic population generalities in the sequences of the galaxies analysed by \citet{Mar17}, \citet{Hal18} 
and \citet{Vul19}. \citet{Hal18} add the varying global environments where galaxies above the CD-19 threshold belong to. In our case, the fact of SF properties not 
only driven by stellar mass (Section~\ref{subsec:dependence}) may be considered a cause too.

Concerning the last remark, Table~\ref{tab:C1} lists the linear regression results in annular comparison sets. Slopes marked with * are the best representatives 
(closest ones from SDTs) of the slopes resulting from compiling both sample data in the corresponding comparison set. Table~\ref{tab:6} lists the samples, either 
perturbed (tA to tJ) or control (initial C), hosting these representatives. Small fonts indicate that the slope is the flattest of the set (see Table~\ref{tab:C1}). 
``f1'' gives the frequencies of best representatives found. ``f2'' follows suit for the flattest slopes in each set. Lastly, ``f3'' tells the frequencies where the 
flattest slope belongs to the perturbed sample. Notice first from Table~\ref{tab:6} that the frequencies of best representatives, f1, can not be ignored either 
annularly (row) or by subsample (column). They represent, at least, 21\,\% (17/80, 40\,\% annulus) and 22\,\% (11/50, SFG LTS subsample) respectively. Similarly, 
from these representatives, the frequencies of being the flattest slope, f2, are far from being unimportant either annularly or by subsample too. They represent 
18\,\% (3/17, 40\,\% annulus) and 32\,\% (8/25, SFG Red subsample) respectively and even more. Finally, from these f2 frequencies, the cases in which the perturbed 
sample is majority, f3, are all but the 20\,\% annulus (4/9), and the SFG ETS (0/12) subsample. We conclude from this all that perturbed galaxies tend to flatten 
the SFMS.

Few best representatives emerge from Fig.~\ref{f5} (5/40). The reason is that the probability of finding the best match gets reduced with the accuracy of the model, 
\textit{e.g.} very low slope errors sometimes of the order of zero (thousandths). Though the frequencies of these representatives are favorable for perturbed galaxies 
(3/5), the few cases prevent us from doing a meaningful analysis. 
\subsection{Central and off-central distinctions}
\label{subsec:centrally}

\citet{BaBa15} compare the EW (H$\alpha$) distributions of CALIFA survey merging and isolated SFGs. Their results are based on a central and an 
extended projected apertures. Their sample distributions in the central aperture exhibit a low likelihood of coming from a common parent distribution. Distributions 
of their extended aperture exhibit the opposite. This contrast proves the SF differences/similitudes along inner/outer extents between merging and isolated objects. 
In contrast, our H$\alpha$ line emission distributions as function of R$_{e}$ (percentage radius/R$_{e}$ fractions, see Table~\ref{tab:3}) suggest that the bigger 
discrepancies between the control and perturbed samples are found within middle radii (40, 60 and 80\,\%). AD and permutation tests on our EW (H$\alpha$) distributions 
result in likelihoods of practically zero along the annular sequence for all subsamples. In addition, from the KS-Peacock two-sample tests in Table~\ref{tab:C1}, 
we obtain annular medians per subsample considering only D$_{\mathrm{2DKS}}$ values marked with *. The resulting trend is similar as that one from Fig.~\ref{f5}, 
\textit{i.e.} higher values distinguish the central 2D distributions whereas lower, rather constant values characterize the off-central ones. 

Furthermore, \citet{Mor15} use smooth particle hydrodynamics to quantify the induced SF extent in galaxy-pair encounters. The merger phase is ignored due to its 
complexity. They find that, whenever enhanced SF is triggered in the nucleus, this is always followed by suppression of activity at larger galactocentric radii. 
We find no evidence of this in our profiles but slight increments in the centres of the perturbed SFG LTS, SFG and all subsamples. Perturbed AGN-like, SFG Red and 
SFG Green show pronounced enhancements and SFG ETS and SFG Blue seem to show no enhancements at all. Notice that important differences characterize both works. 
In the sectioning of the radial extension, \citet{Mor15} use concentric spherical shells of variable widths whereas we use deprojected annuli with widths result 
of comparing fixed fractions of flux. They also integrate the SFRs all along the interaction time scale whereas we give instantaneous-interacting snapshots. However, 
our H$\alpha$ line emission distributions (as function of R$_{e}$) showing the lowest likelihoods of equality within middle radii seem to agree.

\citet{Mor15} also affirm that the interaction time scale must be long enough for the pair configuration to exhibit both enhancement and suppression. According 
to their figure 3, suppression disappears for primary galaxies in post-interacting stages (coalescence and post-coalescence) whilst the SF burst remains. To test 
both, this and the rejuvenation theories, we review the interaction scenarios of perturbed AGN-like galaxies since these are the most favorable ones in the SFR 
properties. SDSS composed fields containing these galaxies are eye-ball reviewed by two members of the author list. To approximate the role at play (primary or 
secondary), absolute magnitudes (\textit{r} and \textit{g} bands) are compared with those of their respective closest companions. After an impartial evaluation 
of each scenario, we find that perturbed AGN-like galaxies are mostly primaries (33/37) but only four appear actually coalescing and only three have signatures 
of post-interaction. For primaries in interaction stages, figure 3 of \citet{Mor15} depicts an enhancement and the slightest suppression. Figures~\ref{f7} and 
\ref{f8} (AGN-like only) illustrate central suppressions (already catalogued as quenching signatures) and then mostly enhancements.

In general, we find signatures of both central and off-central distinctions and also of rejuvenation within our data. 

\section{Summary and conclusions}
\label{sec:conc}

By means of the tidal perturbation parameter, CALIFA survey ELGs are sampled into tidally and non-tidally perturbed (control) objects, the former, on a galaxy-pair 
approach. Ten samples consisting of tidally-perturbed objects match, as proper as possible, the stellar mass and redshift distributions as well as the morphological 
and photometric properties of the control sample. Even the global source of gas excitation is brought to balance. Powerful tools such as IFS and spectral synthesis 
of SPs allow us to obtain spatially-resolved properties for the highly-reliable (deprojected) star-forming regions inhabiting these galaxies. Resolved comparisons 
are conducted at the closest stellar mass densities. Even fairer, AGN-like plus SFG galaxies split up into colours and morphological groups are further distinguished. 
Several effects on SF consequently emerge:

\begin{enumerate}
 \item As distributed on the SFMS plane, most of regions in perturbed galaxies exhibit flatter slopes than those for regions in the control analogues (Fig.~\ref{f5}). Though 
 regions in perturbed objects are indeed older than those in control ones, their offsets with the average SFMS, current-to-past rates and SFR intensities 
 are never inferior (Figs.~\ref{f7} and \ref{f8}). Contrasts are favorable indeed for perturbed AGN-like, SFG Red and SFG Green galaxies. For perturbed SFG LTS, SFG and all galaxies, contrasts 
 are also favorable, though only in the centres. Inside-out quenching signatures are found in AGN-like, SFG Red and all subsamples irrespective of either control 
 or perturbed.
 \item Steeper slopes for regions in perturbed AGN-like, SFG Red and SFG ETS objects do not repeat for regions closest in stellar mass density taken from the rest 
 subsamples (Fig.~\ref{f9}). Dissimilarities in density distributions on the SFMS plane between the former and all types together neither depend nor on the amounts nor on the stellar 
 mass concentration of the regions. Besides, linear fits, either flatter or steeper, do not depend on the stellar mass density of the star-forming regions.
 \item Regions in perturbed AGN-like galaxies tend to be younger and the differences that advantage them in offset, specific rate and intensitiy are notable (Figs.~\ref{f6} to \ref{f8}). A 
 review of the interaction scenarios of these galaxies suggests that they are mostly primaries, however, their signs of suppressed central SF are related to quenching.
 \item For perturbed galaxies, weak but detectable correlations result from relating their integrated properties (SFR, its offset and the sSFR) to their tidal 
 perturbation parameters (Fig.~\ref{f10}). We can give just conjectures regarding this weakness that go from depletions of the gas reservoirs to encounters located not at the 
 pericentre. The correlations may explain the typical SFMS flattening that mostly characterizes the regions in these galaxies. Those exceptions might be explained by a 
 dominant stellar mass, so that no local effect is detected. In sum, SF may also depend on the gravitational torques exerted by the closest companions.
 \item On the SFMS plane, slopes for regions combined from control and perturbed galaxies are regularly best represented by the flatter slopes for regions in perturbed galaxies (Table~\ref{tab:6}). The inclusion of the 
 regions in perturbed objects tends to flatten the SFMS.
\end{enumerate}

On the contrary, \citet{Sch19} find tidal interactions neither influencing total sSFRs nor the scale radius of SF relative to the scale radius of the stellar light. 
However, after discussing all systematics that the estimation of the \textit{f} parameter may involve, they do not discard the possibility of tidal interactions 
enhancing SF in either close pairs or low mass galaxy groups. They propose that either minor mergers or the simple infall of gas from the in-between intergalactic 
medium are real facts in gas-rich galaxy pairs \citep{Jan17}. 

The slight central differences in the SF properties favorable for perturbed galaxies (of SFG LTS, SFG and all types) may reflect contrasts in mass accretion 
rates \citep{Hal18}. Since their SF suppression is less than that at the centres of control objects (of the same types), tidal perturbations may be driving 
gas inflows \citep{Mor15,Ell18b}. We intend to confirm this with a forthcoming analysis of oxygen abundances.
 

\section*{Acknowledgements}
\footnotesize{
Authors wish to thank an anonymous Referee for her/his comments and suggestions that improved this work. A. Morales-Vargas thanks Assistant Editor Bella Lock for her 
kindness.

J. P. Torres-Papaqui and A. Morales-Vargas thank DAIP-UGto for granted support (1006/2016). 

S. F. S\'{a}nchez thanks CONACyT CB-285080, FC-2016-01-1916 and DGAPA-PAPIIT IN100519 projects for their support.

All figures for this paper were possible by the use of \textit{R: A language and environment for statistical 
computing}\footnote{https://www.R-project.org/}. 

The \textsc{starlight}\footnote{http://www.starlight.ufsc.br/} project is supported by the Brazilian agencies CNPq, CAPES and FAPESP and by the 
France-Brazil CAPES/Cofecub program.

The SDSS\footnote{http://www.sdss.org/} is managed by the Astrophysical Research Consortium for the Participating 
Institutions: the Brazilian Participation Group, the Carnegie Institution for Science, Carnegie Mellon University, the Chilean Participation
Group, the French Participation Group, Harvard-Smithsonian centre for Astrophysics, Instituto de Astrof\'{i}sica de Canarias, The Johns Hopkins 
University, Kavli Institute for the Physics and Mathematics of the Universe (IPMU)/University of Tokyo, Lawrence Berkeley National Laboratory, 
Leibniz Institut f\"{u}r Astrophysik Potsdam (AIP), Max-Planck-Institut f\"{u}r Astronomie (MPIA Heidelberg), Max-Planck-Institut f\"{u}r 
Astrophysik (MPA Garching), MaxPlanck-Institut f\"{u}r Extraterrestrische Physik (MPE), National Astronomical Observatories of China, New Mexico 
State University, New York University, Notre Dame University, Observat\'{o}rio Nacional/MCTI, Ohio State University, Pennsylvania State 
University, Shanghai Astronomical Observatory, United Kingdom Participation Group, Universidad Nacional Aut\'{o}noma de M\'{e}xico, University 
of Arizona, University of Colorado Boulder, University of Oxford, University of Portsmouth, University of Utah, University of Virginia, University 
of Washington, University of Wisconsin, Vanderbilt University, and Yale University.

The Calar Alto Legacy Integral Field Area survey is the first legacy survey being performed at Calar Alto and is managed by the CALIFA 
survey Collaboration\footnote{https://bit.ly/2lnB4Um}. All them would like to thank the IAA-CSIC and MPIA-MPG as major partners of the 
observatory, and CAHA itself, for the unique access to telescope time and support in manpower and infrastructures. The CALIFA survey 
Collaboration thanks also the CAHA staff for the dedication to the project.

\textexclamdown\,Gracias Mateo\,!}


\newpage
\section*{DATA AVAILABILITY}
The data underlying this article will be shared on reasonable request to the corresponding author.





\bsp	


\appendix

\section{The tidally-perturbed samples}
\label{sec:app1}

Fundamental and other properties for control and perturbed objects (see Table~\ref{tab:A1}). The latter are the closest, in the five fundamentals (M$_{*}$, \textit{z}, 
morphological group, galaxy colour and dominant excitation source), to each single control object.

\onecolumn
\scriptsize{
\setlength{\tabcolsep}{0.75\tabcolsep}
\begin{longtable}{c@{\hspace{\tabcolsep}}c@{\hspace{\tabcolsep}}c@{\hspace{\tabcolsep}}c@{\hspace{\tabcolsep}}c@{\hspace{\tabcolsep}}c@{\hspace{\tabcolsep}}c@{\hspace{\tabcolsep}}c@{\hspace{\tabcolsep}}c@{\hspace{\tabcolsep}}c@{\hspace{\tabcolsep}}c@{\hspace{\tabcolsep}}c@{\hspace{\tabcolsep}}c@{\hspace{\tabcolsep}}c@{\hspace{\tabcolsep}}c}
 \caption{\scriptsize{Fundamental properties for control (\textit{f}$<-4.5$) and perturbed (\textit{f}$\geq-4.5$) objects. Some other properties are also listed.}}
 \label{tab:A1}\\
\hline                                                                                                                                                           \\
      &Global                                           &                                       &Morpho-                                    &                                         &                                        &Dominant                                 &                                               &                                         &                                              &                                            &H$\alpha$ flux                              &Median                                  &Number                                     &Global                                                                      \\
      &M$_{*}$                                          &                                       &logy                                       &Morpho-                                  &Galaxy                                  &excitation                               &Surface                                        &                                         &                                              &                                            &60\,\%                                      &SP age                                  &of star-                                   &SFR                                                                         \\
 NED  &log$_{10}$                                       &                                       &(Collabo-                                  &logical                                  &colour                                  &source                                   &scale                                          &Axis                                     &\textit{f}                                    &Depro-                                      &radius                                      &log$_{10}$                              &forming                                    &log$_{10}$                                                                  \\
 name &(M$_{\odot}$)\footnotesize{{\color{blue}$^{a}$}} &$z$\footnotesize{{\color{blue}$^{b}$}} &ration)\footnotesize{{\color{blue}$^{c}$}} &group\footnotesize{{\color{blue}$^{d}$}} &type\footnotesize{{\color{blue}$^{e}$}} &(BPT)\footnotesize{{\color{blue}$^{f}$}} &(kpc$^{2}$)\footnotesize{{\color{blue}$^{g}$}} &ratio\footnotesize{{\color{blue}$^{h}$}} &parameter.\footnotesize{{\color{blue}$^{i}$}} &jection.\footnotesize{{\color{blue}$^{j}$}} &/R$_{e}$\footnotesize{{\color{blue}$^{k}$}} &(yr)\footnotesize{{\color{blue}$^{l}$}} &regions\footnotesize{{\color{blue}$^{m}$}} &($\mathrm{M}_{\odot}\,\mathrm{yr}^{-1}$)\footnotesize{{\color{blue}$^{n}$}} \\[1ex]
\hline                                                                                                                                                           \\
\endhead
              &                &        &         &    &      &      &       &     &                 &  &     &                &               &                 \\
\cline{1-4}\\
\multicolumn{14}{l}{\footnotesize{{\color{blue}$^{a}$}}\scriptsize{ Global stellar mass (see Section~\ref{subsec:M_est}).}}\\
\multicolumn{14}{l}{\footnotesize{{\color{blue}$^{b}$}}\scriptsize{ Raw redshift measured from the CALIFA cube.}}\\
\multicolumn{14}{l}{\footnotesize{{\color{blue}$^{c}$}}\scriptsize{ Reclassification of morphologies (see Section~\ref{subsec:gal-col}).}}\\
\multicolumn{14}{l}{\footnotesize{{\color{blue}$^{d}$}}\scriptsize{ Sbc and later types as Late Type Spirals (LTSs). All the rest as Early Type Spirals (ETSs) (which include a few Early Types, ETs).}}\\
\multicolumn{14}{l}{\footnotesize{{\color{blue}$^{e}$}}\scriptsize{ CMD (see Section~\ref{subsec:gal-col} and Fig.~\ref{f3}).}}\\
\multicolumn{14}{l}{\footnotesize{{\color{blue}$^{f}$}}\scriptsize{ [\ion{O}{iii}]/H$\beta$ against [\ion{N}{ii}]/H$\alpha$ \citep[BPT diagnostic diagram,][]{BPT81}.}}\\
\multicolumn{14}{l}{\footnotesize{{\color{blue}$^{g}$}}\scriptsize{ Equivalence of 1\,arcsec$^{2}$.}}\\
\multicolumn{14}{l}{\footnotesize{{\color{blue}$^{h}$}}\scriptsize{ Semiminor to semimajor axis ratio from light moment analyses (see W-14). Also $cos\,\phi$, \textit{i.e.}, the cosine of the inclination angle (see Fig.~\ref{f1}).}}\\
\multicolumn{14}{l}{\footnotesize{{\color{blue}$^{i}$}}\scriptsize{ Tidal perturbation parameter (see Section~\ref{subsec:f}).}}\\
\multicolumn{14}{l}{\footnotesize{{\color{blue}$^{j}$}}\scriptsize{ Deprojection: Yes (Y), no (N). Relocation of each spaxel coordinates on a disk plane.}}\\
\multicolumn{14}{l}{\footnotesize{{\color{blue}$^{k}$}}\scriptsize{ H$\alpha$ emission distribution: radius containing 60\,\% of the H$\alpha$ flux as a function of R$_{e}$ (see Section~\ref{subsec:annuli&profiles}).}}\\
\multicolumn{14}{l}{\footnotesize{{\color{blue}$^{l}$}}\scriptsize{ Median SP age (see Section~\ref{subsec:M_est}).}}\\
\multicolumn{14}{l}{\footnotesize{{\color{blue}$^{m}$}}\scriptsize{ Number of star-forming regions (see Section~\ref{subsec:s-f_rsub}).}}\\
\multicolumn{14}{l}{\footnotesize{{\color{blue}$^{n}$}}\scriptsize{ Global SFR (see Section~\ref{subsec:s-f_rsub}).}}\\
\endfoot
                                                            \multicolumn{15}{c}{Control objects (62)}                                                            \\
\end{longtable}
}              
\twocolumn     
\normalsize{   

\section{An important note on the annular profiles}
\label{sec:app2}

To minimize the influence of stellar mass, in each annular set of spaxel properties, the perturbed spaxels from each trial (tA to tJ, see Section~\ref{subsec:gaxsamp}) 
are paired with control ones by minimizing their differences in $\mathrm{\Sigma}_{*}$. The medians of these annular differences are plotted as the bars in the 
$\mathrm{\Sigma}_{*}$ profiles of Fig.~\ref{fB1}. All profiles of Section~\ref{sec:res} plot each annular difference of the respective property in the same way as the 
median differences of Fig.~\ref{fB1}. Notice, for the $\Sigma_{*}$, that control and perturbed sample profiles overlap since all bar heights are $\sim$0 dex (rounded 
numbers at two figures). Besides, from comparing each annular distribution pair, an important fraction of the AD-permutation tests gives high likelihoods. Cases in which 
at least one likelihood (either AD or permutation) does not reach the half are identified in Figs.~\ref{fB1} and \ref{fB2}. The reason 
why is the unbalance in the number of spaxels or star-forming regions to be paired (see Section~\ref{sec:val}). Such unbalances reduce the probability of finding the best 
minimum absolute difference between paired spaxels. To review this issue, we further show the involved distributions and perform on them the Mann-Whitney (MW) test. The 
null hypothesis of this test is that the distributions to be compared differ by a location shift (mainly the median). Notice that the third numbers, \textit{i.e.} the 
MW p-values in the identified cases (AGN-like, SFG Red and SFG ETS panels of Fig.~\ref{fB1}), are much greater than the statistical level so the null hypothesis is rejected. Moreover, by 
looking at the distributions of Fig.~\ref{fB2}, the only dissimilar ones are those for the central annulus. As suggested by the MW test, the medians and also the IQRs 
are alike except the distribution tails, specifically, those of the control sample (20\,\% annulus). In sum, results for the centres of AGN-like, SFG Red and SFG ETS 
galaxies are the only ones that must be taken with care (both AD and permutation tests do not suggest similitude).

\begin{figure}\centering
   \mbox{\includegraphics[width=\columnwidth]{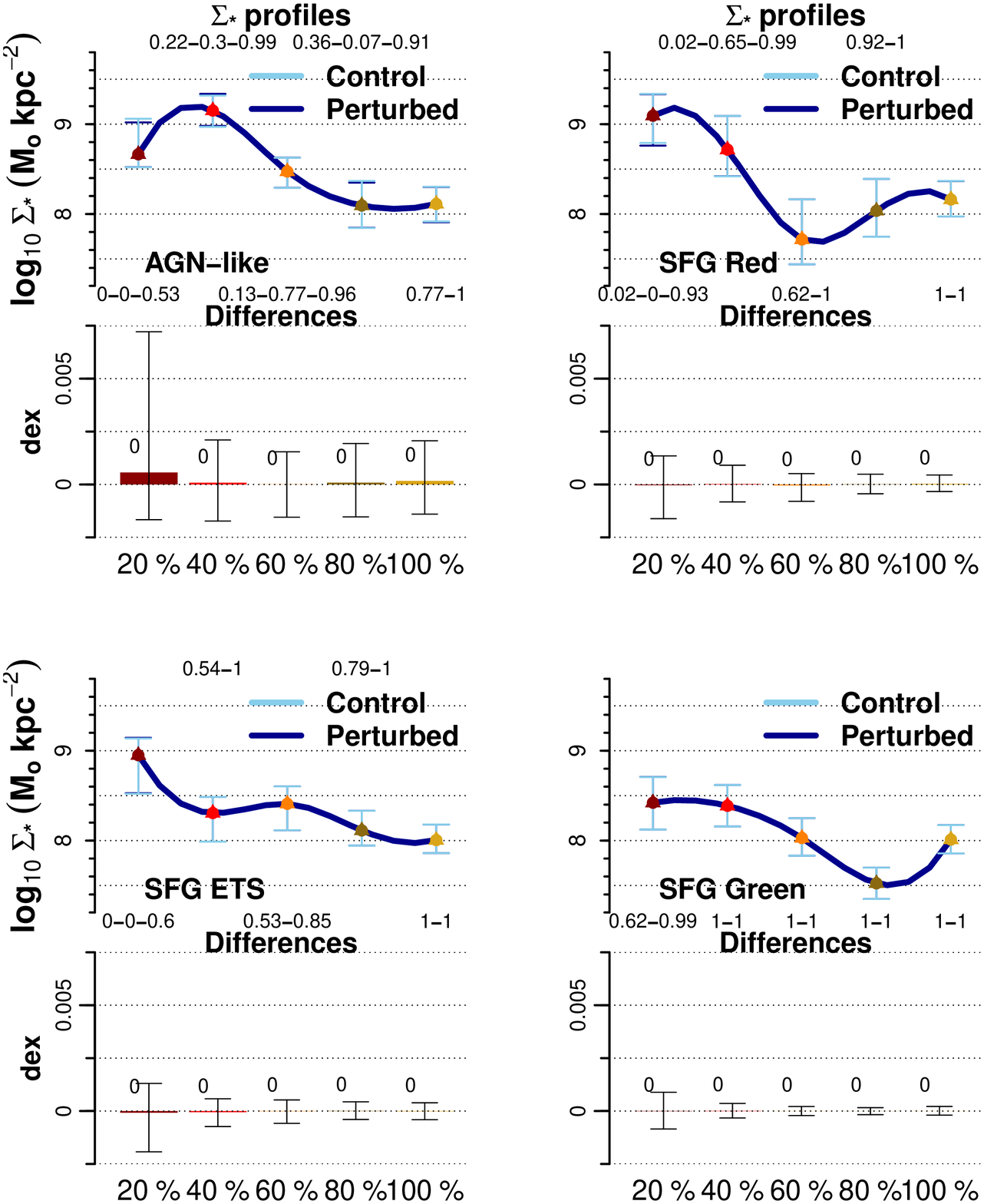}}
   \mbox{\includegraphics[width=\columnwidth]{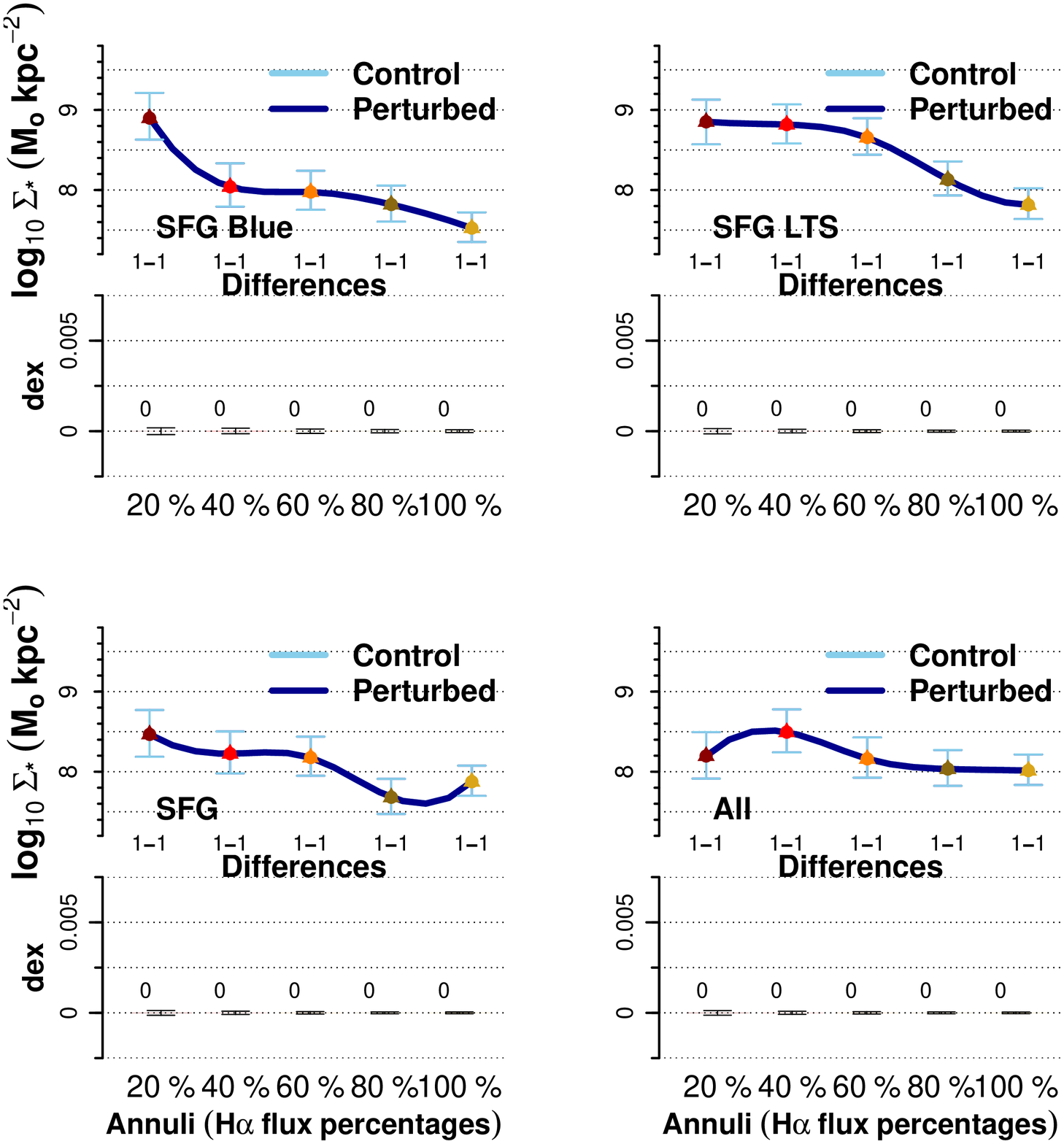}}
\caption{\scriptsize{Annular profiles: $\mathrm{\Sigma}_{*}$. Throughout this work, all profiles use the amounts of annular star-forming regions per sample and subsample as
in Fig.~\ref{f5}. Five consecutive-outward annuli denote the radial extension (see Section~\ref{subsec:annuli&profiles}). ``Differences'' (bar heights by always 
subtracting the control values from the perturbed ones) are the medians of the annular distributions of differences (differences by pairing sample spaxels which 
are the closest in $\mathrm{\Sigma_{*}}$). Bar lines depict the interquartile ranges (IQRs, 1st to 3rd) of the distributions of differences. Symbols are both sample 
values giving each Difference. Symbol lines depict the IQRs of the annular distributions of each sample. Above and below the profiles, likelihoods from AD-permutation 
tests are found for each pair of sample distributions. Finally, a third statistics, Mann-Whitney (MW) test \emph{p-value}, is given when at least one of the other 
two statistics results in a value lower than 0.5 (see text).}}
   \label{fB1} 
\end{figure}

\begin{figure}\centering
   \mbox{\includegraphics[width=\columnwidth]{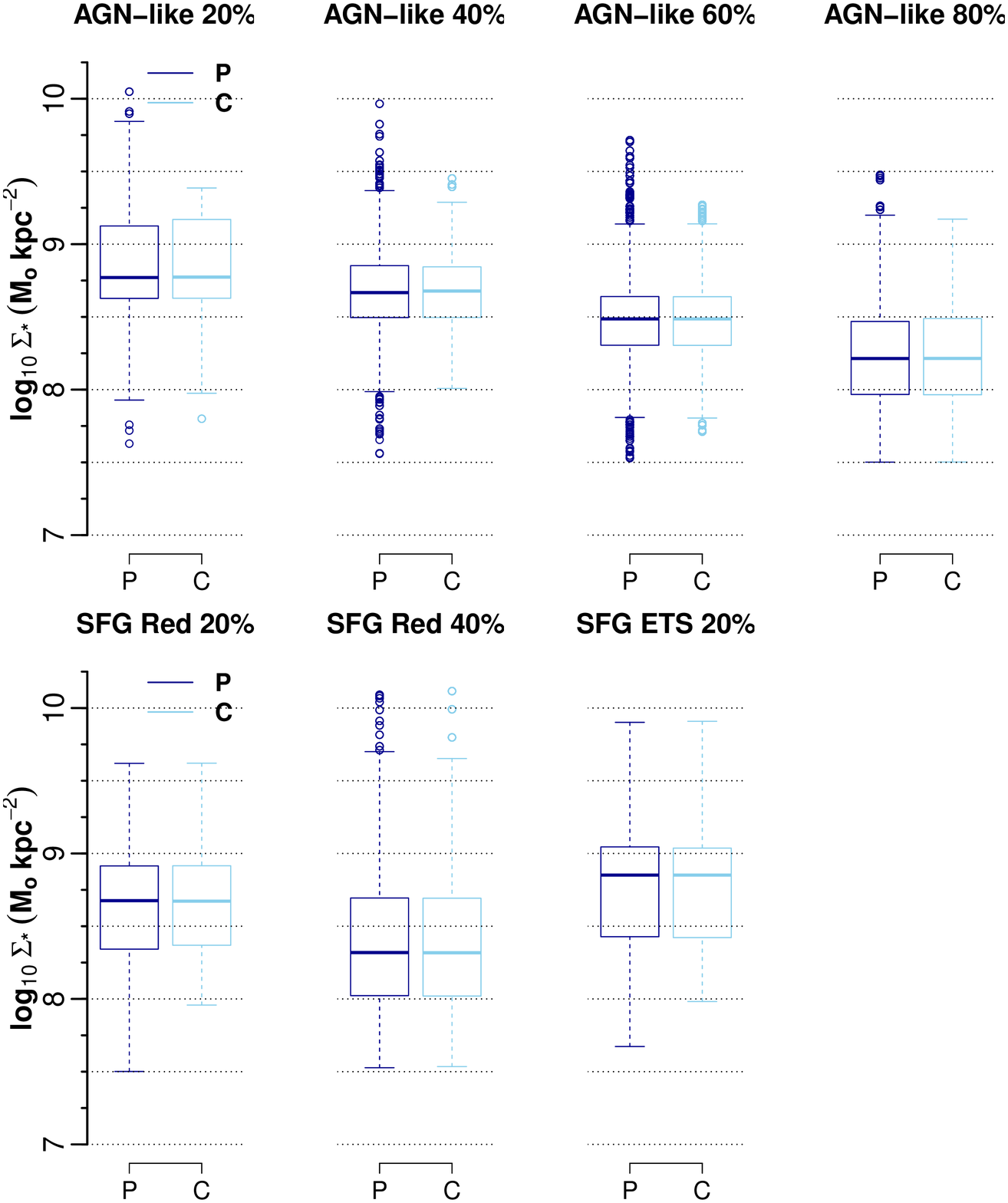}}
\caption{\scriptsize{$\mathrm{\Sigma}_{*}$ distributions (P, perturbed, against C, control) for the annular cases of Fig.~\ref{fB1} in which at least one 
of the statistical tests results in a likelihood lower than 0.5.}}
   \label{fB2} 
\end{figure}

\newpage
\section{Resolved SFMSs}
\label{sec:app3}

The annular comparison sets are shown in Table~\ref{tab:C1}. It  lists the linear regression coefficients of control and tA to tJ samples all on the SFMS plane. In each set, slopes 
marked with * are the closest ones to the slopes which result from the data in the respective set. Only significance difference tests (SDTs, see Section~\ref{subsec:mssf_1}) 
with values above the statistical level are considered for this and that slope with the highest test value is marked (see Section~\ref{subsec:representatives}). 
The criterion is that slopes are suggested to be different when the SDT value falls below the statistical level. Only these cases of different slopes are considered 
for comparisons (see Section~\ref{subsec:mssf_1}). At last, from the K-S/Peacock two-sample test, $\mathrm{D_{2DKS}}$ differences marked with * reject the null 
hypothesis of a parent distribution as the origin. 

\onecolumn
\scriptsize{
\setlength{\tabcolsep}{0.325\tabcolsep}
\begin{longtable}{r@{\hspace{3\tabcolsep}}c@{\hspace{0.075\tabcolsep}}c@{\hspace{3\tabcolsep}}c@{\hspace{0.075\tabcolsep}}c@{\hspace{3\tabcolsep}}c@{\hspace{0.075\tabcolsep}}c@{\hspace{3\tabcolsep}}c@{\hspace{0.075\tabcolsep}}c@{\hspace{3\tabcolsep}}c@{\hspace{0.075\tabcolsep}}c@{\hspace{3\tabcolsep}}c@{\hspace{0.075\tabcolsep}}c@{\hspace{3\tabcolsep}}c@{\hspace{0.075\tabcolsep}}c@{\hspace{3\tabcolsep}}c@{\hspace{0.075\tabcolsep}}c@{\hspace{3\tabcolsep}}c@{\hspace{0.075\tabcolsep}}c@{\hspace{3\tabcolsep}}c@{\hspace{0.075\tabcolsep}}c}
 \caption{\scriptsize{Comparison annular sets: control (C) and perturbed galaxies (tA-tJ, see text) on the SFMS plane. Linear regression slope (\textit{s}) and 
intercept (\textit{i}, M$_{\odot}$\,yr$^{-1}$\,kpc$^{-2}$) both with standard error (\textit{e}). All models are statistically significant but two for control 
galaxies in the AGN-like sets (20\,\% annulus, 3rd and 4th from left to right). Slopes with * are the closest to the ones if combining both sample data. 
Significance difference tests (SDTs) per pair of slopes are shown too. $\mathrm{D_{2DKS}}$ differences (the maximum absolute ones between cumulative probability 
distributions, \citealt{Pea83}, \citealt{Yua17}) with * suggest that both distribution functions come from different parent ones. Finally, numbers of paired 
star-forming regions with annular totals at the very left (see Fig.~\ref{f5}).}}
 \label{tab:C1}\\                 
\hline                                                                                                                                                           \\
                   &C                 &tA                 &C                  &tB                &C                 &tC                 &C                  &tD                &C                 &tE                 &C                  &tF                 &C                 &tG                 &C                  &tH                &C                 &tI                 &C                  &tJ                \\[1ex]
\hline                                                                                                                                                           \\
\endhead
	                    &\multicolumn{20}{c}{\textbf{AGN-like}}                                                                                                                                                                                                                                                                                                                                                                \\[1ex]
\textbf{20\,\%} \textit{s}  &\phantom{$-$}0.24 &\phantom{$-$}1.18  &\phantom{$-$}0.43  &\phantom{$-$}1.16 &\phantom{$-$}0.10 &\phantom{$-$}0.99  &$-$0.12            &\phantom{$-$}1.38*&\phantom{$-$}0.29 &\phantom{$-$}1.02  &\phantom{$-$}0.24  &\phantom{$-$}1.29  &\phantom{$-$}0.53 &\phantom{$-$}1.45  &\phantom{$-$}0.25  &\phantom{$-$}1.19 &\phantom{$-$}0.33 &\phantom{$-$}1.14  &\phantom{$-$}0.31  &\phantom{$-$}1.00 \\
         \textit{e}         &\phantom{$-$}0.09 &\phantom{$-$}0.07  &\phantom{$-$}0.11  &\phantom{$-$}0.09 &\phantom{$-$}0.14 &\phantom{$-$}0.10  &\phantom{$-$}0.22  &\phantom{$-$}0.12 &\phantom{$-$}0.07 &\phantom{$-$}0.06  &\phantom{$-$}0.08  &\phantom{$-$}0.06  &\phantom{$-$}0.15 &\phantom{$-$}0.15  &\phantom{$-$}0.08  &\phantom{$-$}0.07 &\phantom{$-$}0.08 &\phantom{$-$}0.06  &\phantom{$-$}0.06  &\phantom{$-$}0.05 \\
         \textit{i}         &$-$3.61           &$-$12.00           &$-$5.31            &$-$11.77          &$-$2.34           &$-$10.22           &$-$0.36            &$-$13.79          &$-$4.05           &$-$10.45           &$-$3.63            &$-$12.85           &$-$6.20           &$-$14.33           &$-$3.70            &$-$12.11          &$-$4.44           &$-$11.57           &$-$4.20            &$-$10.44          \\
\end{longtable}
}              
\twocolumn     
\normalsize{   


\label{lastpage}
\end{document}